\documentclass[aps,prb,english,floatfix,reprint,twocolumn,showpacs,superscriptaddress,longbibliography,10pt,tightenlines]{revtex4-1} 

\usepackage[utf8]{inputenc}
\usepackage{amssymb,amsmath}
\usepackage{graphicx}
\usepackage{natbib}
\usepackage{multirow}
\usepackage{braket}
\usepackage{color}
\usepackage{subcaption}
\usepackage{ragged2e}
\DeclareCaptionJustification{justified}{\justifying}
\usepackage[final]{hyperref} % adds hyper links inside the generated pdf file
\hypersetup{
	colorlinks=true,       % false: boxed links; true: colored links
	linkcolor=red,        % color of internal links
	citecolor=red,        % color of links to bibliography
	filecolor=magenta,     % color of file links
	urlcolor=blue         
}

\begin{document}

\title{Quantum quenches in an interacting field theory: full quantum evolution vs. semi-classical approximations}

\author{D. Sz{\'a}sz-Schagrin}
\affiliation{BME Momentum Statistical Field Theory Research Group,\\Department of Theoretical Physics, Budapest University of Technology and Economics}
\author{I. Lovas}
\affiliation{Department of Physics and Institute for Advanced Study,
Technical University of Munich}
\affiliation{ Munich Center for Quantum Science and Technology (MCQST)}
\author{G. Tak{\'a}cs}
\affiliation{BME Momentum Statistical Field Theory Research Group,\\Department of Theoretical Physics, Budapest University of Technology and Economics}
\affiliation{MTA-BME Quantum Correlations Group (ELKH), Department of Theoretical Physics, Budapest University of Technology and Economics}
\date{3rd October 2021}

\begin{abstract}
We develop a truncated Hamiltonian method to investigate the dynamics of the $(1+1)d~\phi^4$ theory following quantum quenches. The results are compared to two different semi-classical approaches, the self-consistent Gaussian approximation and the truncated Wigner approximation, and used to determine the range of validity of these widely used approaches. We show that the  self-consistent approximation is strongly limited in comparison to the truncated Hamiltonian method which for larger cutoffs is practically exact for the parameter range studied. We find that the self-consistent approximation is only valid when the effective mass is in the vicinity of the renormalised mass. Similarly to the self-consistent approximation, the truncated Wigner approximation is not able to capture the correct mass renormalisation, and breaks down for strong enough interactions where the bare mass becomes negative. We attribute the failure of TWA to the presence of a classical symmetry broken fixed point. Besides establishing the truncated Hamiltonian approach as a powerful tool for studying the dynamics of the $\phi^4$ model, our results on the limitation of semi-classical approximations are expected to be relevant for modelling the dynamics of other quantum field theories.
\end{abstract}

\maketitle

\section{Introduction}

Interacting quantum many-body systems are in the focus of interest of contemporary physics. Due to the growing number of experimental realisation of isolated quantum many-body systems, particularly in cold atomic experiments, the interest in study of out-of-equilibrium dynamics of such systems greatly  increased\cite{Hofferberth_2007,Meinhert_2013,Langen_2013,Trotzky_2012,Gring_2012,Fukuhara_2013,Cheneau_2012,Kukuljan_2019,Sotiriadis_2010}. A paradigmatic and experimentally realistic protocol for initiating out-of-equilibrium dynamics is the quantum quench. In this scenario the system is initially in equilibrium, such as a thermal state or ground state of some pre-quench Hamiltonian. At $t = 0$ some parameter of the Hamiltonian is suddenly changed, pushing the system out of equilibrium and leading to a non-trivial time evolution.

For interacting field theories, the time evolution generally can not be solved exactly, making necessary the use of suitable approximation methods. In this work we consider two important non-perturbative semi-classical approaches: the (Gaussian) self-consistent approximation (SCA)~\cite{Sotiriadis_2010, Essler2019, Essler2021}, and the truncated Wigner approximation (TWA)~\cite{Polkovnikov2003,POLKOVNIKOV2010}. In the SCA approach the interaction is taken into account in a mean-field approximation, leading to a self-consistency condition. For the $\phi^4$ theory it was first developed by Sotiriadis and Cardy~\cite{Sotiriadis_2010} to study the time evolution of the effective mass and extract thermal characteristics of the system. The TWA relies on the classical equations of motion to approximate the dynamics, while the quantum fluctuations are incorporated in the fluctuating initial state. 

Both the SCA and TWA are valuable tools in studying the dynamics of interacting field theories, however, they are uncontrolled approximations, warranting careful testing against other methods. The family of truncated Hamiltonian approaches provides a more controlled way to investigate non-equilibrium dynamics in quantum field theories, taking into account the full quantum dynamics\cite{Rakovszky_2016,2017PhLB..771..539H,2018ScPP....5...27H,2018PhRvL.121k0402K,2019JHEP...08..047H,2019PhRvA.100a3613H,2021PhRvD.104b1702K}. The first application of Hamiltonian truncation to quantum field theories was the truncated conformal space approach (TCSA) to numerically study relevant perturbations of conformal field theories, introduced by Yurov and Zamolodchikov~\cite{Yurov_1989}. Several variants were subsequently developed to treat a larger class of models\cite{1991IJMPA...6.4557Y,Feverati_1998,Fonseca_2001,2015PhRvD..91b5005H,Rychkov_2015,Rychkov_2016,Bajnok_2016}, as well as boundary\cite{1998NuPhB.525..641D,2001NuPhB.614..405B} and defect problems\cite{2014NuPhB.886...93B}; for a recent review c.f. the paper by James et al.\cite{2018RPPh...81d6002J} Although applications of Hamiltonian truncation are mostly dominated by 1+1 dimensional field theories, it is possible to extend the approach to higher dimensions as well \cite{2015PhRvD..91b5005H}. 

Here we use a truncated Hamiltonian approach (THA) built upon a massive free boson\cite{2015PhRvD..91b5005H,Rychkov_2015,Rychkov_2016,Bajnok_2016}, and develop it further to describe out-of-equilibrium dynamics following quantum quenches in the $(1+1)d$ $\phi^4$ theory. We demonstrate its accuracy and efficiency, and then proceed to compare its results to those of the SCA and TWA approaches.

The structure of the paper is as follows. Section \ref{sec:THA} introduces the $1+1$ dimensional $\phi^4$ theory, and specifies the family of quantum quenches and the observables considered. In Section \ref{sec:THA} we also present the truncated Hamiltonian approach and demonstrate that for a large range of parameters it gives a numerically very accurate result for the time evolution. We then proceed to consider the semi-classical approximations: Section \ref{sec:SC} is devoted to the self-consistent, while Section \ref{sec:twa} turns to the truncated Wigner approximation, comparing their results to the one obtained from THA. Section \ref{sec:conclusions} contains the summary and discussion of the results.

\section{Quantum quenches and the truncated Hamiltonian method}\label{sec:THA}

The $(1+1)d$ $\phi^4$ model in finite volume $L$ is given by the Hamiltonian
\begin{equation}\label{eq:hamilton}
H = H_{\text{KG}}^m + \frac{\lambda}{4!}V_4,
\end{equation}
where 
\begin{equation}\label{eq:HKGm}
H_{\text{KG}}^m=\int\limits_0^L dx :\left(\frac{1}{2}\Pi^2+\frac{1}{2}(\partial_x\phi)^2+\frac{m^2}{2}\phi^2\right):
\end{equation}
is the Klein-Gordon Hamiltonian with mass $m$, the scalar field $\phi$ and its conjugate momentum $\Pi$ satisfy the canonical commutation relations
\begin{equation}\label{eq:CCR}
    \left[\phi(t,x),\Pi(t,y)\right]=i\delta(x-y)\,,
\end{equation}
and 
\begin{equation}\label{eq:defVn}
    V_n = \int_0^L :\phi^n:dx\,.
\end{equation}
We use periodic boundary conditions $\phi(t,x + L) = \phi(t,x)$, and the semicolons denote the normal ordering with respect to the free boson modes corresponding to the mass $m$. Throughout this work we confine ourselves to the $\mathbb{Z}_2$-symmetric phase $m^2>0$. We use units in which $m = 1$, and so the volume can be parameterised the dimensionless parameter $l = mL$, while the quartic coupling $\lambda$ is measured in units of $m^2$.

A general quantum quench in the $\phi^4$ theory corresponds to changing $m_0 \rightarrow m_1$ and $\lambda_0 \rightarrow \lambda_1$ at the initial time $t=0$,  so the Hamiltonian before the quench is
\begin{equation}\label{eq:prequench_ham}
H_0 = H_{\text{KG}}^{m_0} + \frac{\lambda_0}{4!}V_4\,,
\end{equation}
while the post-quench is governed by
\begin{equation}\label{eq:postquench_ham}
H_1 = H_{\text{KG}}^{m_1} + \frac{\lambda_1}{4!}V_4\,.
\end{equation}
The initial state is taken to be the vacuum state of $H_0$: $H_0\ket{\psi_0} = E_0\ket{\psi_0}$ with the minimum possible eigenvalue $E_0$. The $t>0$ time evolution is unitary and governed by $H_1$: 
\begin{equation}
\ket{\psi(t)} = e^{-i H_1 t}\ket{\psi_0}\,.
\end{equation}
In this work we set the initial coupling $\lambda_0=0$, which means that the initial state can be constructed exactly since it is the ground state of a free massive boson. 

The time-evolution of an observable $\mathcal{O}$ is given by
\begin{equation}
\braket{\mathcal{O}} := \bra{\psi(t)}\mathcal{O}\ket{\psi(t)}\,.
\end{equation}
In the present work we focus on the two-point function $C_2(x, t)$
\begin{equation}
C_2(x, t) = \braket{ :\phi(x,t)\phi(0,t): }\label{expvalue}
\end{equation}
and its Fourier-components $\braket{:\phi_k(t)\phi_{-k}(t):}$. When the post-quench coupling $\lambda_1$ is also zero, the evaluation of the time evolution becomes a simple exercise using the Bogolyubov transformation relating the eigenmodes of the pre-quench and post-quench Hamiltonians. Due to the periodic boundary conditions, the allowed Fourier modes are given as $k = 2\pi n/l$ with $n\in\mathbb{Z}$, and the normal ordering in Eq. \eqref{eq:defVn} is defined relative to the modes corresponding to the post-quench mass $m_1$. 

The finite volume formulation is a necessary requirement for the truncated Hamiltonian method, and results in a discrete spectrum. In the next step a UV energy cutoff is introduced which restricts the Hilbert space to have a finite dimension, where the Hamiltonian and the observables can be represented as finite matrices. We remark that using light-cone quantisation it is possible to formulate a truncated Hamiltonian approach directly in infinite volume \cite{2016JHEP...07..140K,2020arXiv200513544A}.

Here we work in the eigenbasis of the Klein-Gordon Hamiltonian of mass $m = 1$ in a finite volume parameterised by the dimensionless variable $l=mL$. In principle, this mass unit can be different from both the pre-quench and post-quench masses $m_0$ and $m_1$, but in practice it is convenient to set  $m=m_1$. Since both the initial state and the post-quench Hamiltonian are translationally invariant, the space of states can be restricted to the zero-momentum subspace. Then the energy cutoff $\Lambda$ can be parameterised as
\begin{equation}
\frac{\Lambda}{m} = 4 \pi n_\text{max} / l
\end{equation}
where $n_\text{max}$ is the quantum number of the maximum momentum mode that can be excited in the limit of small $m$. Keeping only the low-energy eigenstates as basis, the finite matrices of $V_n$ and eventually expectation values such as the Fourier modes of the two-point function
\begin{equation}
C_2(n,t)=\braket{\phi_k(t)\phi_{-k}(t)} \qquad k=\frac{2\pi n}{l}
\end{equation}
can be computed by simple algebra\cite{Rychkov_2016,Bajnok_2016}. These results however depend on the value of $\Lambda$ and differ from the exact values, with the deviation called the truncation error. For the $\phi^4$ theory, since the quartic interaction is a relevant perturbation of the Klein-Gordon field, the truncation error tends to zero as the cutoff is taken to infinity. However, increasing the cutoff can be computationally expensive, and so further procedures might be necessary to reduce the truncation error. The simplest prescription is leading order renormalisation group improvement\cite{Rychkov_2015}, which results in local counter terms summing up the high energy contributions and defining an effective Hamiltonian which eliminates the dominant part of the truncation effects. Using the notation introduced in Eq. \eqref{eq:defVn}, the renormalised effective Hamiltonian is\cite{Rychkov_2015}
\begin{equation}\label{eq:renormHam}
H_\text{eff} = H_{\text{KG}}^{m} + g_4 V_4 + \kappa_0 V_0 + \kappa_2 V_2 + \kappa_4 V_4
\end{equation}
where the $\kappa_i$ (to leading order) are given by\cite{Rychkov_2015} 
\begin{equation}
\begin{split}
&\kappa_0 = - {\int^\infty_{\Lambda}} \frac{dE}{E-\epsilon_*}[g_4^2\mu_{440}(E) + g_2^2\mu_{220}(E)]\\
&\kappa_2 = - {\int^\infty_{\Lambda}}  \frac{dE}{E-\epsilon_*}[g_4^2\mu_{442}(E) + g_2g_4\mu_{422}(E)]\\
&\kappa_4 = - {\int^\infty_{\Lambda}}  \frac{dE}{E-\epsilon_*}g_4^2\mu_{444}(E)
\end{split}
\end{equation}
where $g_n$ is the coupling of the $V_n$ term at infinite cutoff in the perturbation added to the free massive Hamiltonian $H_{\text{KG}}^{m}$, i.e.
\begin{equation}
    g_2=0\,,\quad g_4=\frac{\lambda}{4!}\,,
\end{equation}
and the functions determining the renormalisation group running of the couplings are
\begin{equation}
\begin{split}
& \mu_{220}(E) = \frac{1}{\pi E^2} \quad \mu_{422} =\frac{12}{\pi E^2} \quad \mu_{444}(E) = \frac{36}{\pi E^2}\\
&\mu_{440}(E) = \frac{1}{E^2}\left[\frac{18}{\pi^3}(\log E/m)^2-\frac{3}{2\pi}\right]\\
&\mu_{442}(E) = \frac{72\log E/m}{\pi^2E^2} \,,
\end{split}
\end{equation}
while $\epsilon_*$ is a reference energy\cite{Rychkov_2015}, which we set to zero. 

\begin{figure}
\begin{center}
\includegraphics[width=\linewidth]{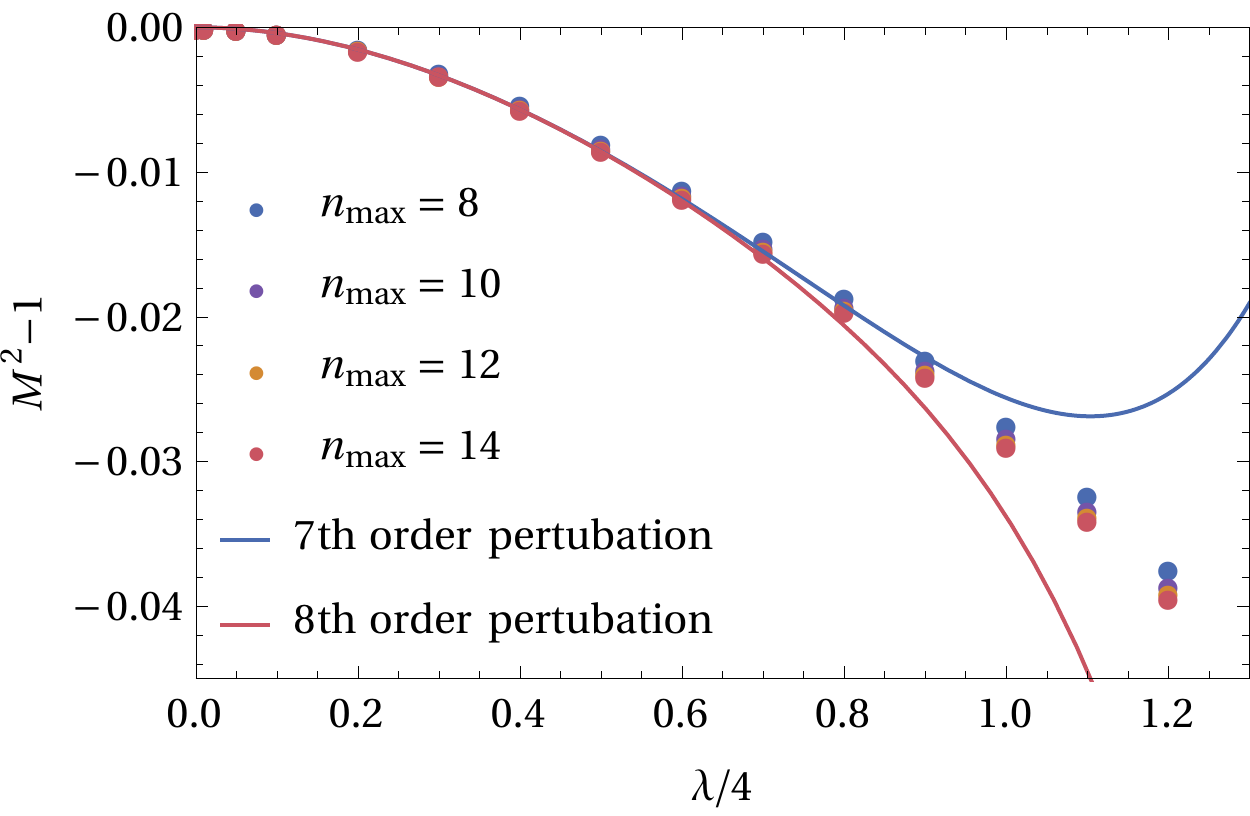}
\caption{Dependence of the gap of the spectrum on the scalar self-coupling $\lambda$ computed in $l=10$ volume in the perturbative regime using units $m=1$, for different values of the cutoff parameter $n_\text{max}$, including the leading order RG improvement Eq. \eqref{eq:renormHam}. The 7th and the 8th order perturbative results\cite{Serone_2018} are shown with solid lines.}
\label{fig:gap}
\end{center}
\end{figure}

The accuracy of the spectrum of the renormalised Hamiltonian can be tested by comparing the mass gap obtained from THA to perturbative results\cite{Serone_2018}. These results also illustrate the efficiency of the leading order RG improvement \eqref{eq:renormHam} to eliminate cutoff dependence\cite{Rychkov_2015}.

For the time evolution we use the post-quench mass $m_1$ to define our units, i.e. we set $m=m_1$. The initial state can be represented using its exact expression in terms of the free bosonic modes with the post-quench mass by means of Bogolyubov transformation. However, its normalisation must be changed from the exact value to ensure that the initial vector has unit norm on the truncated Hilbert space. 

\begin{figure*}
\begin{center}
\begin{subfigure}[b]{0.48\textwidth}
\includegraphics[width=\columnwidth]{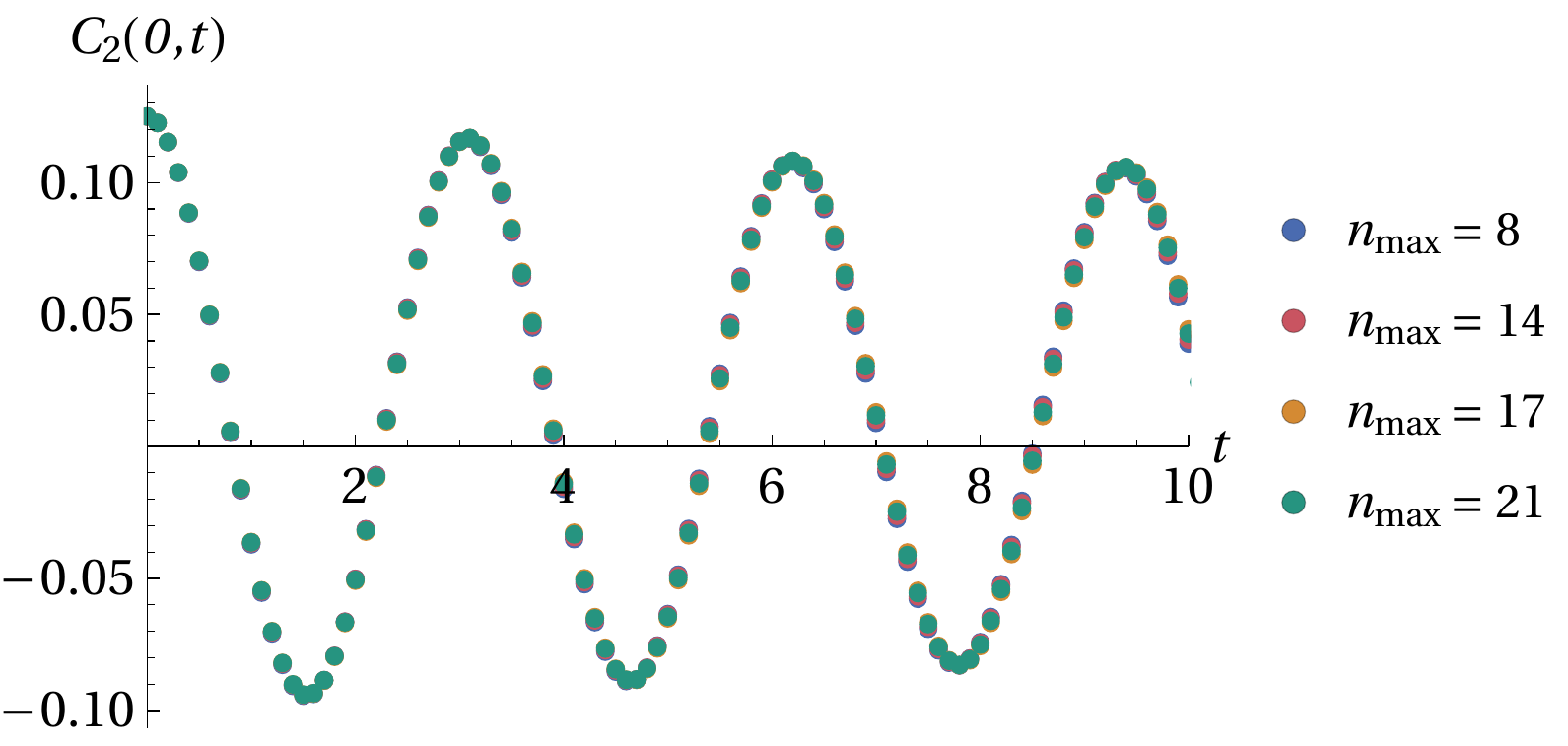}
\caption{Quench $(m_0,\lambda_0) = (0.8,0) \rightarrow (m_1,\lambda_1) = (1,2)$}
\end{subfigure}
\begin{subfigure}[b]{0.48\textwidth}
\includegraphics[width=\columnwidth]{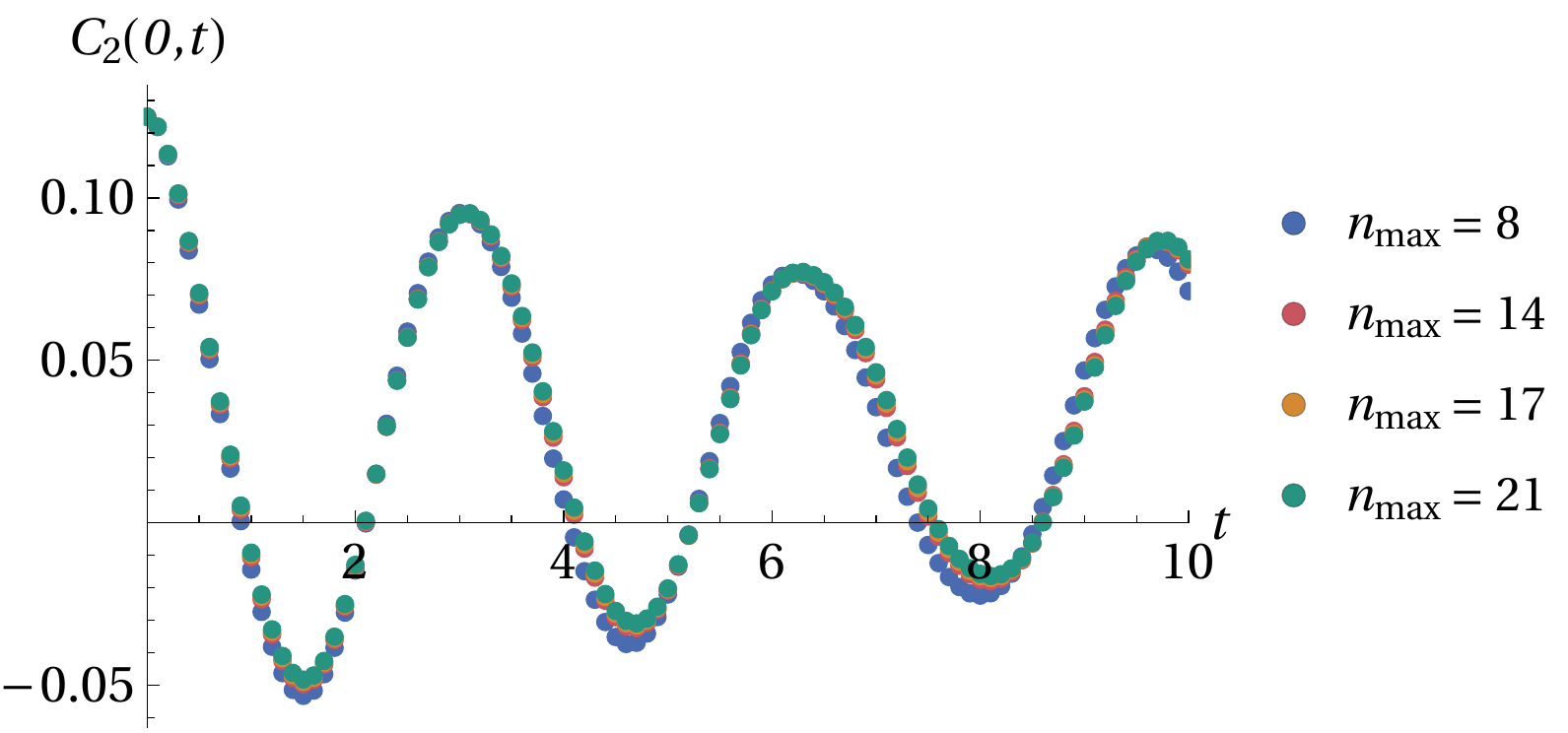}
\caption{Quench $(m_0,\lambda_0) = (0.8,0) \rightarrow (m_1,\lambda_1) = (1,8)$}
\end{subfigure}
\caption{Time evolution of the zero-mode of the two-point function $C_2(n=0,t)$ for (a) a relatively small and (b) a much larger quench quench for different values of the cutoff as computed by THA. Different values of $n_\text{max}$ correspond to different colours: $n_\text{max} = 8$, $14$, $17$, and $21$ corresponding to truncated state spaces of dimension cc.~$400$, $16000$, $90000$ and $700000$ states respectively.}
\label{fig:tha-truncation}
\end{center}
\end{figure*}

The cutoff dependence of the THA evolution is shown in Fig. \ref{fig:tha-truncation}. Note that even for relatively large quenches the cutoff dependence is negligible. The results obtained using the renormalised Hamiltonian essentially coincide with the ones at the highest cutoff $n_\text{max} = 21$. As a result, the THA results for the time evolution can be considered essentially exact for the quenches used in Sections \ref{sec:SC} and \ref{sec:twa}. Therefore using them as a basis for comparison gives a direct test of the applicability and accuracy of semi-classical methods.

We also note that the time evolution for $t\gtrsim L$ is expected to deviate from the result in infinite volume due to finite size effects resulting from excitations travelling around the spatial circle. Therefore we restrict all our simulations to times $t\leq L$; although obtaining results for larger times is not a problem per se, they would differ from the dynamics in thermodynamic limit and be affected by the periodic boundary conditions.

\section{Self-consistent approximation vs. truncated Hamiltonian approach}\label{sec:SC}

A simple method for approximating (\ref{expvalue}) is the self-consistent (a.k.a. mean-field) approximation which was developed by Cardy and Sotiriadis\cite{Sotiriadis_2010}. 

In the self-consistent approximation the quartic coupling in (\ref{eq:postquench_ham}) is replaced with a mean field approximation:
\begin{equation}\label{hf}
\phi^4 ~\rightarrow~ 6\braket{\phi^2}\phi^2 -3\braket{\phi^2}^2\,,
\end{equation}
which is equivalent to dropping the fully connected part of the quartic interaction term $:\phi^4:$. Since the second term is just a constant shift in the energy, it does not affect the time evolution and therefore it can also be dropped. The first term corresponds to a shift in the particle mass and therefore the dynamics is essentially governed Klein-Gordon Hamiltonian with a time-dependent mass. The  mass itself can be computed from the two-point correlation function in a self-consistent way. Naively the time dependent effective mass $m_{\text{eff}}(t)$ can be defined by the relation
\begin{equation}\label{gap-eq}
m_{\text{eff}}^2(t) = m_1^2 + \frac{\lambda}{2}\sum_k\braket{\phi_k(t)\phi_{-k}(t)}\,.
\end{equation}
Since the effective mass also appears in $\braket{\phi^2(t)}$ the right hand side also depends on $m_{\text{eff}}(t)$, equation \eqref{gap-eq} is a self-consistency condition determining $m_{\text{eff}}(t)$. However, due to ultraviolet divergences the actual value of the effective mass must be defined through a renormalisation procedure, resulting in the gap equation
\begin{equation}\label{gap-eq-renorm}
m_{\text{eff}}^2(t) = m_1^2 + \frac{\lambda}{2}\sum_k\left(\braket{\phi_k(t)\phi_{-k}(t)} -\frac{1}{2\omega_k}\right)
\end{equation}
where $\omega_k = \sqrt{m_1^2 + k^2}$. This renormalisation prescription is equivalent to normal ordering the product $\phi_k\phi_{-k}$ with respect to $m_1$, and matches the normal ordering used in the truncated Hamiltonian approach. In our explicit calculations we also set the mass scale $m$ which defines our units equal to the post-quench mass $m_1$, which results in conventions matching those in the THA approach, making the comparison between the results of the THA and SCA straightforward.

The time evolution of the correlator $\braket{\phi_k\phi_{-k}}$ is obtained by solving the equation of motion for the field $\phi$ in the mean field approximation \eqref{hf}, which is complemented by the self-consistent evaluation of the  effective mass from \eqref{gap-eq-renorm}. This procedure can be implemented by discretising the $(t,k)$ space and solving the equations by iterative application of the following steps\cite{Sotiriadis_2010}:
\begin{enumerate}
\item[1.] Construct a time-dependent mode frequency function $\Omega_k(t)$ for every mode $k$ from
\begin{equation}\label{omega-eq2}
\frac{\ddot\Omega_k}{2\Omega_k} - \frac{3}{4}\left(\frac{\dot\Omega_k}{\Omega_k}\right)^2 + \Omega_k^2 = \omega_k(t)^2\,.
\end{equation}
with the initial condition
\begin{equation}\label{initcond2}
\Omega_k(0) = \omega_k(0)\quad\quad \dot\Omega_k(0) = 0\,.
\end{equation}
where 
\begin{equation}\label{k-dep}
\omega_k(t) = \sqrt{ m_{\text{eff}}^2(t) + k^2}\,.
\end{equation}
\item[2.] Compute $\braket{\phi_k(t)\phi_{-k}(t)}$ for every $k$ from
\begin{equation}\label{corr-k}
\begin{split}
&\braket{\phi_k(t)\phi_{-k}(t)} = \frac{1}{\Omega_k(t)}\bigg[\frac{\omega_k(0)^2 + \omega_{0k}^2}{2\omega_k(0)\omega_{0k}} +\\
&\frac{\omega_k(0)^2 - \omega_{0k}^2}{2\omega_k(0)\omega_{0k}}\cos\left(2\int_0^tdt'\Omega_k(t')\right)\bigg]\,,
\end{split}
\end{equation}
where $\omega_{0k}^2 = m_0^2 + k^2$.

\item[3.] Obtain $m_{\text{eff}}^2(t)$ from the gap equation
\begin{equation}\label{gap-eq-renorm2}
m_{\text{eff}}^2(t) = m_1^2 + \frac{\lambda_1}{2}\sum_k\left(\braket{\phi_k(t)\phi_{-k}(t)} -\frac{1}{2\omega_k(0)}\right)\,.
\end{equation}
\item[4.] Move to the next time step $t \rightarrow t + dt$.
\end{enumerate}

\begin{figure*}
\begin{center}
\begin{subfigure}[b]{\textwidth}
\includegraphics[width=0.48\textwidth]{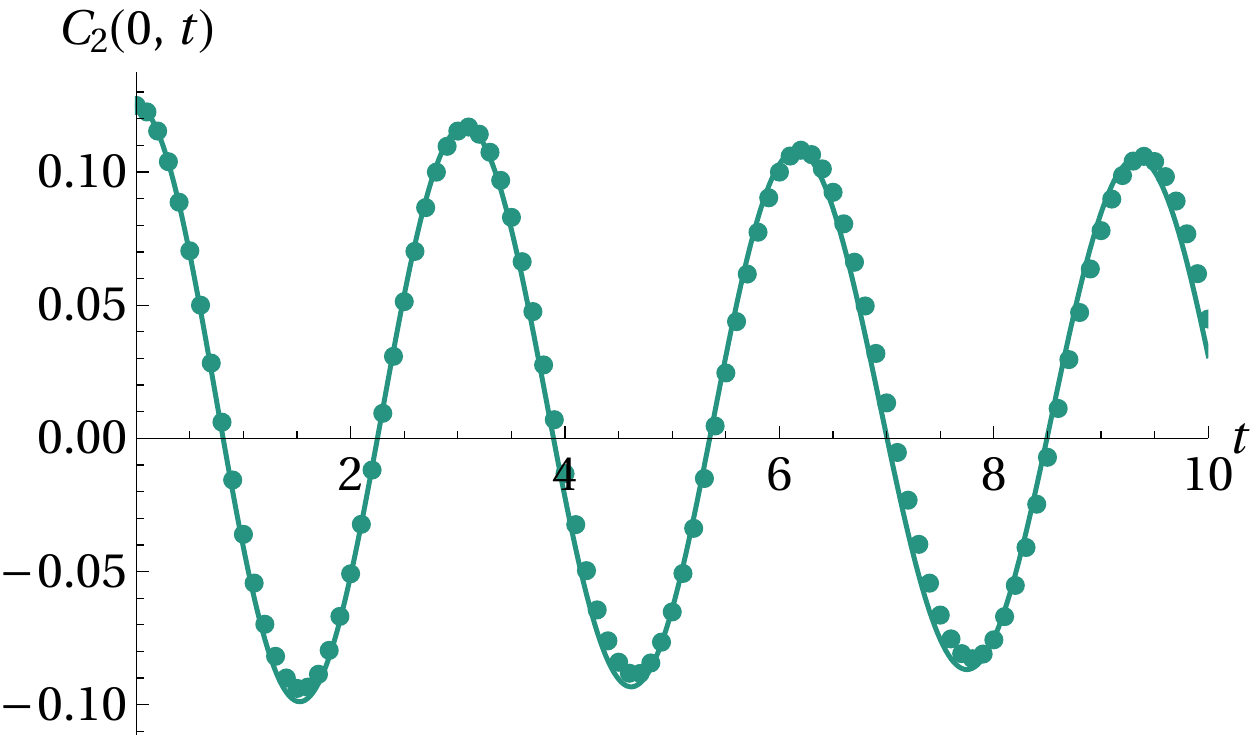}
\includegraphics[width=0.48\textwidth]{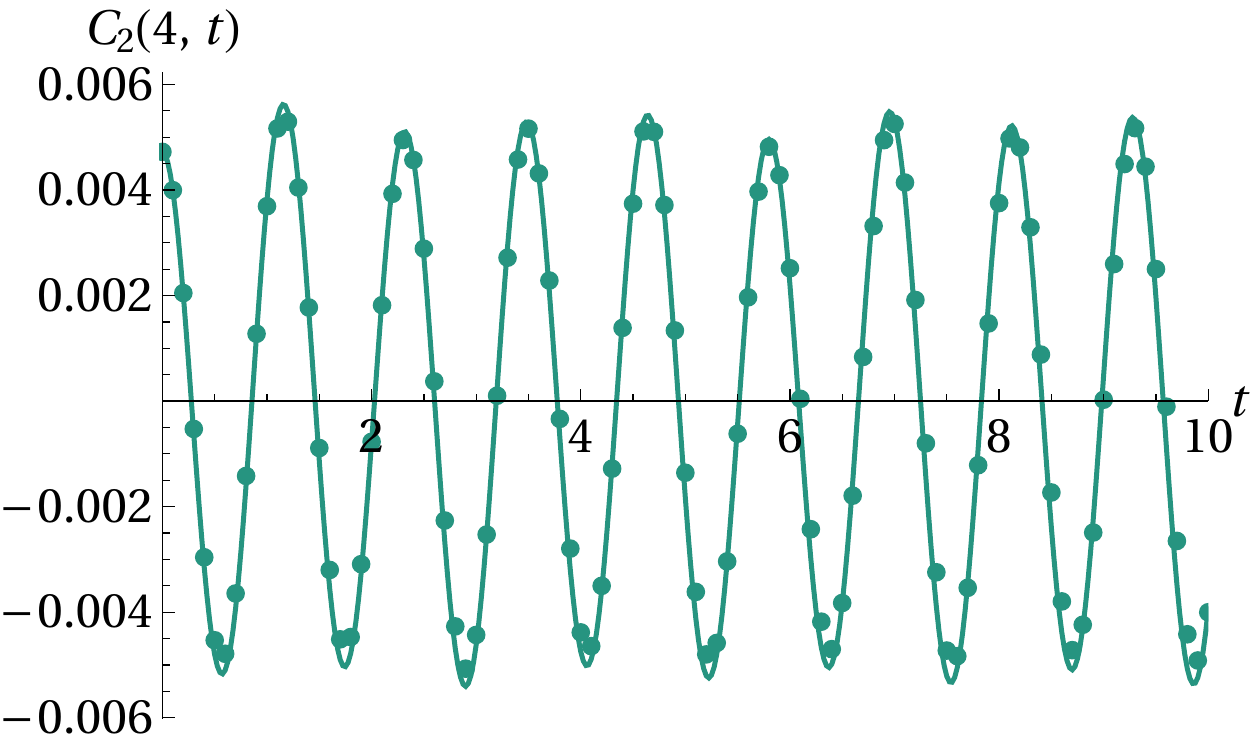}
\caption{$(m_0,\lambda_0) = (0.8,0) \rightarrow (m_1,\lambda_1) = (1,2)$}
\label{fig:c2-lambda2}
\end{subfigure}
\begin{subfigure}[b]{\textwidth}
\includegraphics[width=0.48\textwidth]{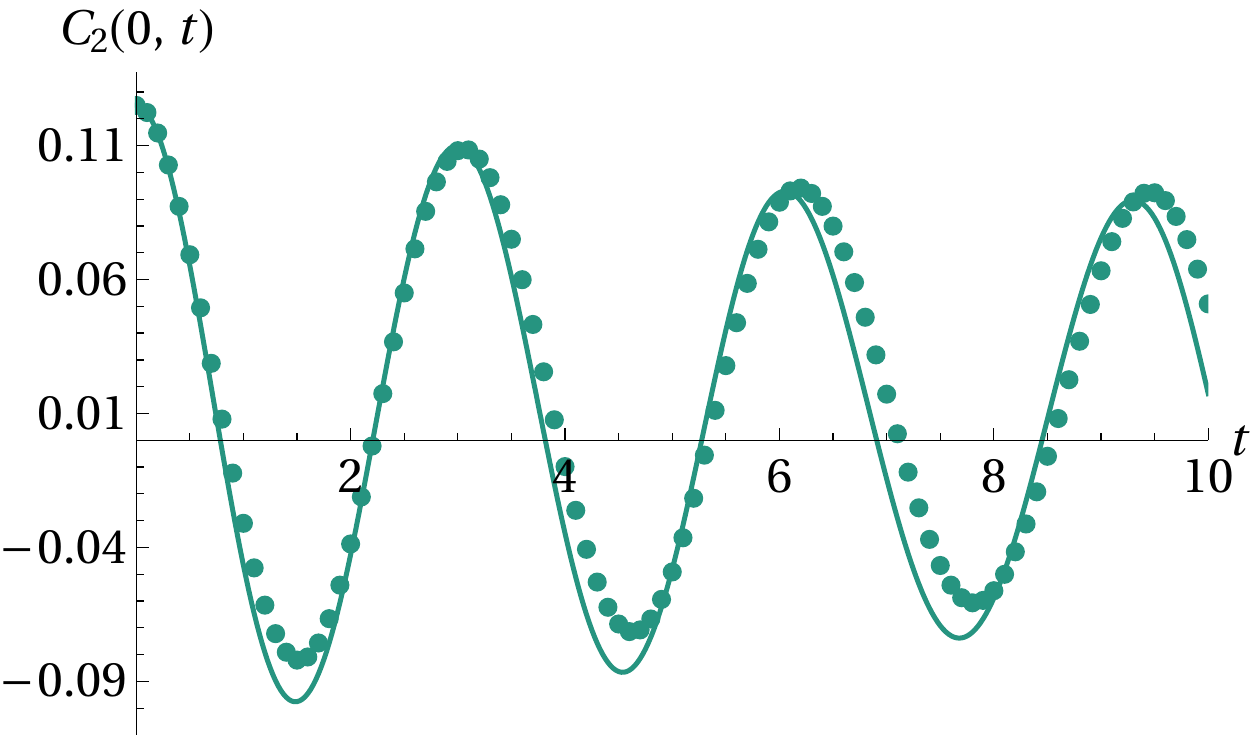}
\includegraphics[width=0.48\textwidth]{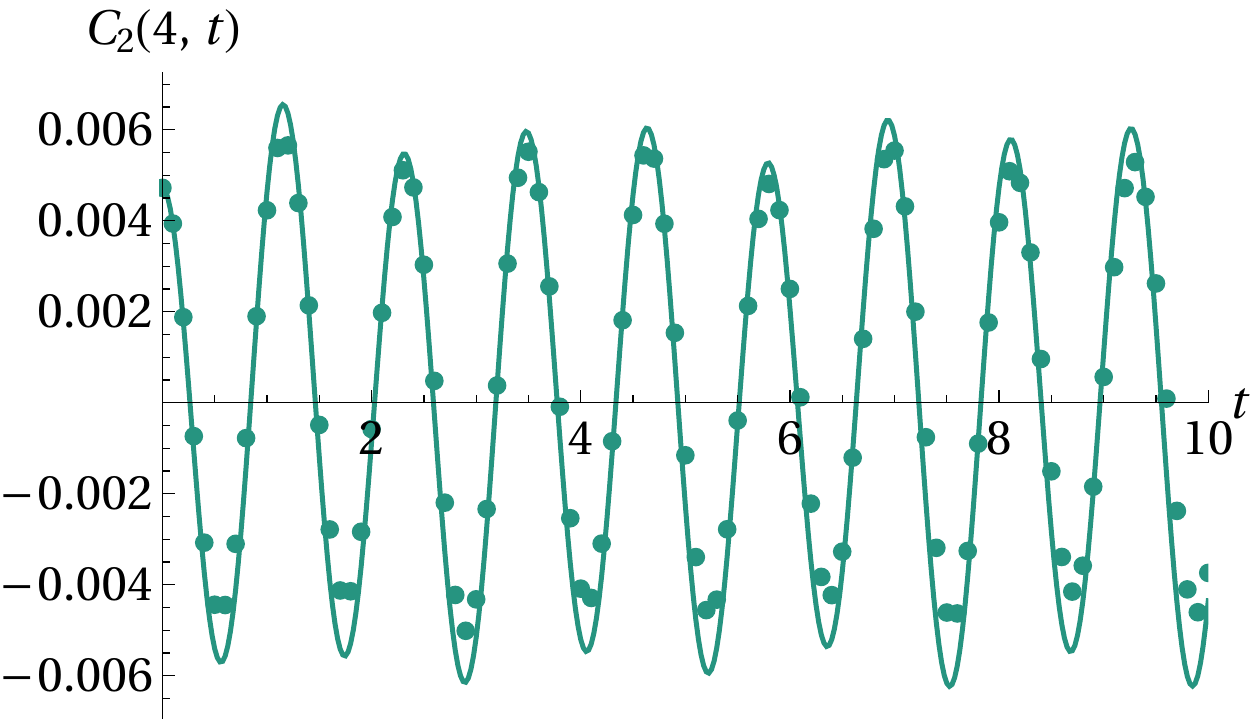}
\caption{$(m_0,\lambda_0) = (0.8,0) \rightarrow (m_1,\lambda_1) = (1,4)$}
\label{fig:c2-lambda4}
\end{subfigure}
\begin{subfigure}[b]{\textwidth}
\includegraphics[width=0.48\textwidth]{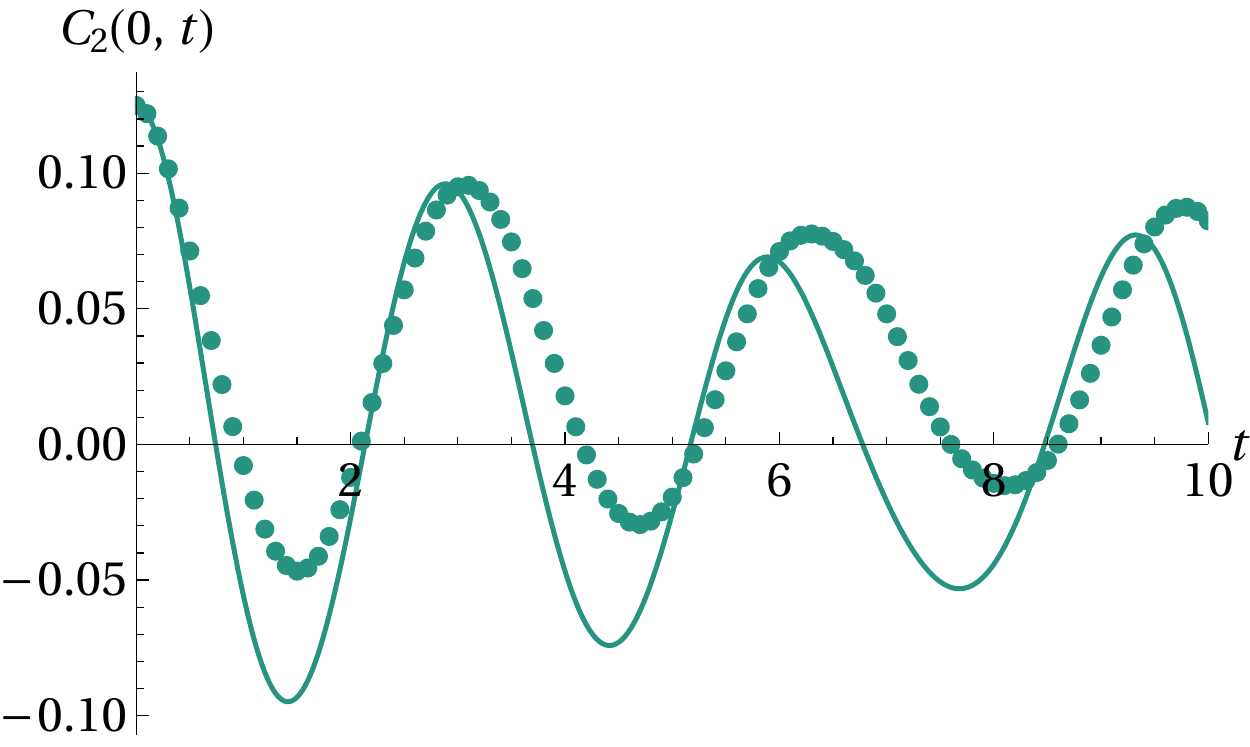}
\includegraphics[width=0.48\textwidth]{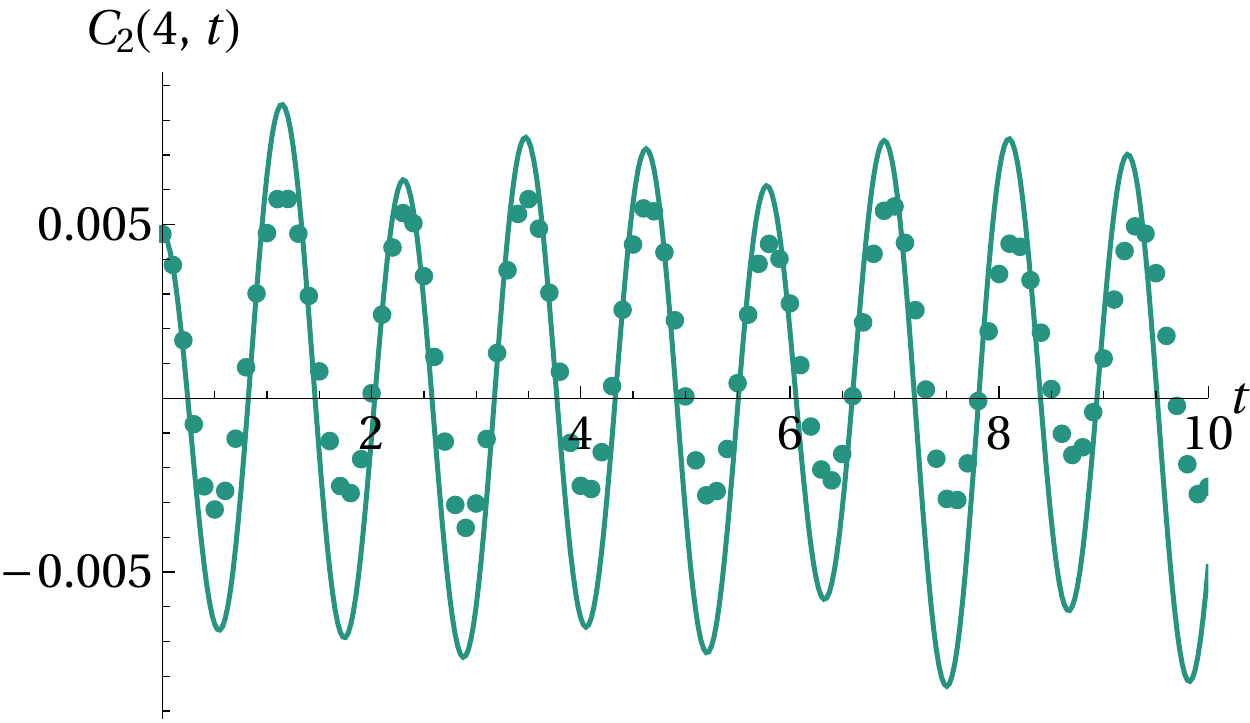}
\caption{$(m_0,\lambda_0) = (0.8,0) \rightarrow (m_1,\lambda_1) = (1,8)$}
\label{fig:c2-lambda8}
\end{subfigure}
\caption{Time evolution of the $n = 0$ and $n = 4$ mode of $C_2$ for quenches with the mass changed from $0.8$ to $1$, with different post-quench interactions. The dots denote the numerical data from the truncated Hamiltonian approach at the cutoff  $n_\textrm{max}=21$ computed using the renormalisation improvement \eqref{eq:renormHam}, while the solid line is the result of the self-consistent approximation.}
\label{fig:c2-lambdas}
\end{center}
\end{figure*}

Since we compare the results to THA evolution in a finite volume $L$ (with $l=mL=10$), the momentum is naturally discretised as $k=2\pi n/L$ with $n\in\mathbb{Z}$, and the set of momenta for which the time evolution of $\braket{\phi_k\phi_{-k}}$ is constructed is made finite by introducing an upper cutoff $|n|\leq N_\textrm{max}$. The numerical results converge fast when increasing $N_\textrm{max}$; however, choosing it excessively large makes the procedure unstable and must be avoided. An additional parameter is the time step $dt$ which must be set small enough so that the results become independent of its choice. The numerical procedures were validated by simulating mass quenches with $\lambda_1=0$ for which the mean-field approximation is exact, and the truncated Hamiltonian approach was found to yield practically exact results. 

\begin{figure*}
\vspace{2mm}
\begin{center}
\begin{subfigure}[b]{0.48\textwidth}
\includegraphics[width=\columnwidth]{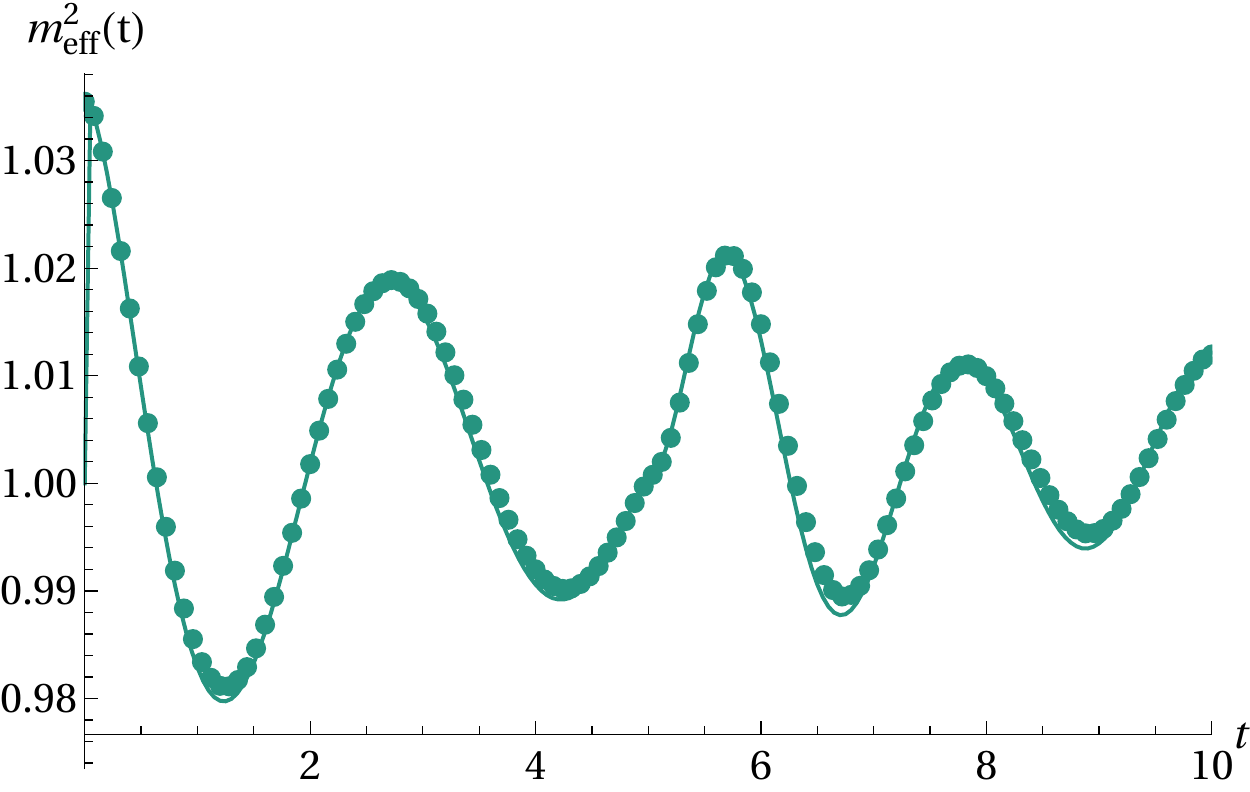}
\caption{Quench $(m_0,\lambda_0) = (0.8,0) \rightarrow (m_1,\lambda_1) = (1,2)$}
\end{subfigure}
\begin{subfigure}[b]{0.48\textwidth}
\includegraphics[width=\columnwidth]{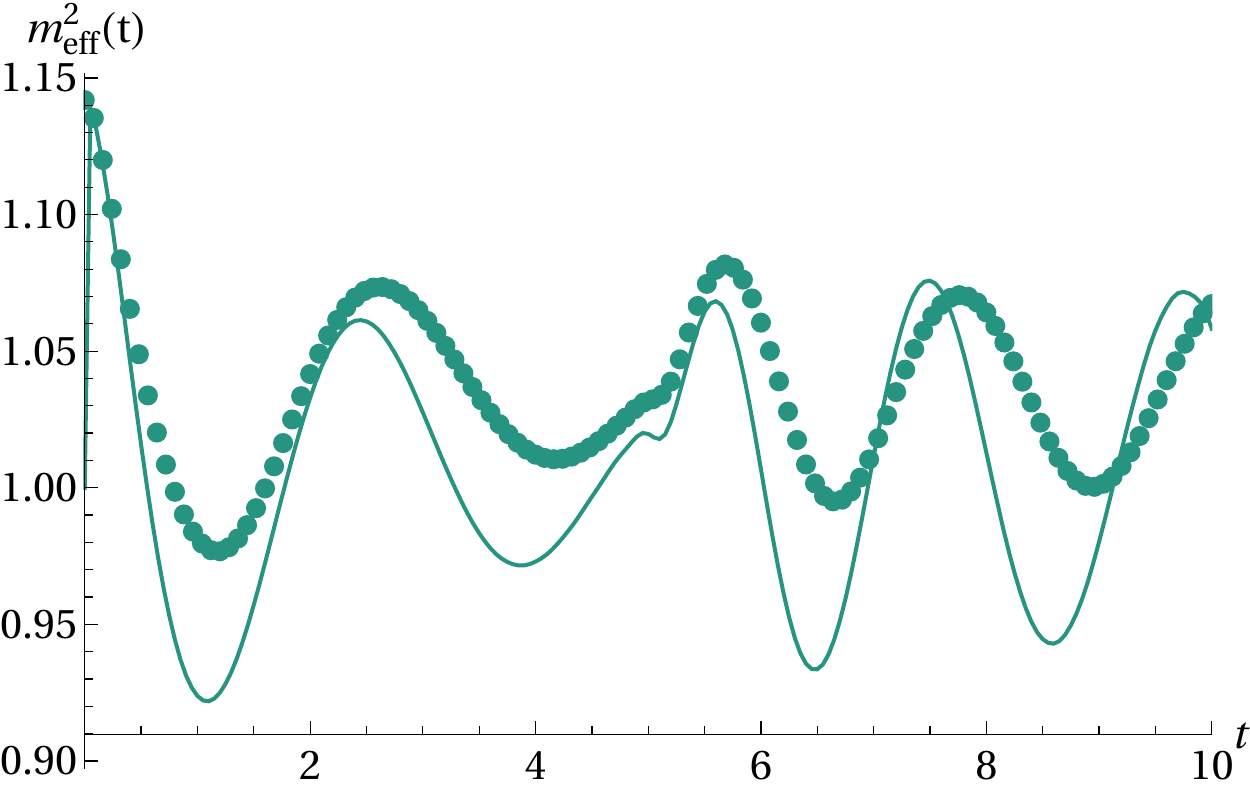}
\caption{Quench $(m_0,\lambda_0) = (0.8,0) \rightarrow (m_1,\lambda_1) = (1,8)$}
\end{subfigure}
\caption{The effective mass in the THA and SCA approximations. The markers denote the truncated Hamiltonian results, while the solid line shows the self-consistent solution corresponding to the largest value of $n_\text{max}$ is shown. For the smaller quench (coupling is in the perturbative regime), the agreement between the results is clear. We can also see that the effective mass is in the close vicinity of 1 (top). The difference for the larger quench is much bigger, which can be understood as the effective mass getting further from 1.}
\label{fig:meff}
\end{center}
\end{figure*}

The results for interacting quenches are illustrated in Fig. \ref{fig:c2-lambdas}, in which we fixed the change in the mass parameter and varied the post-quench coupling $\lambda_1$. We show the zero wave number component $C_2(0,t)$ and also a higher wave number component $C_2(4,t)$ for the correlator. The results show that the self-consistent approximation becomes progressively less reliable as the coupling increases. Looking at the component $C_2(0,t)$, it is clear that a main source of the discrepancy is that the SCA does not reproduce the frequencies present in the time evolution and this discrepancy increases with the coupling. There is clearly a discrepancy in the amplitudes as well:  Eq. \eqref{corr-k} implies that this is linked to the discrepancy in the frequencies. 

The data at the strongest coupling in Fig.~\ref{fig:c2-lambda8} also show that for higher modes (illustrated by $C_2(4,t)$) the frequencies of the two methods match well, suggesting that the disagreement stems from the difference in the relevant mass scale of the two approaches. This follows from the simple observation that for an effective mass $m_{\rm eff}$, the frequency associated with number $k$ is
\begin{equation}
\Omega_k \sim \sqrt{m_{\rm eff}^2 + k^2}\,.
\end{equation}
The mass scale $m_{\rm eff}$ thus determines the low-frequency part of the spectrum, but the frequencies of higher modes are dominated by the contribution of the wave number $k$, so using the precise scale $m_{\rm eff}$ becomes less important.
Indeed, the two approaches lead to a different effective mass $m_{\rm eff}$. For the THA, $m_\text{eff}$ can be determined by substituting the numerically evaluated correlations into \eqref{gap-eq-renorm2}, while for the SCA it is obtained using \eqref{gap-eq-renorm2} as part of the self-consistent time evolution. As shown in Fig. \ref{fig:meff}, the two effective masses agree well for small coupling, while show a strong discrepancy for larger coupling. This leads to a pronounced difference in the frequencies of low energy modes, whereas the temporal frequencies of higher are expected to agree better, in accordance with Fig.~\ref{fig:c2-lambda8}.

\begin{figure}
\vspace{2mm}
\begin{center}
\includegraphics[width=\columnwidth]{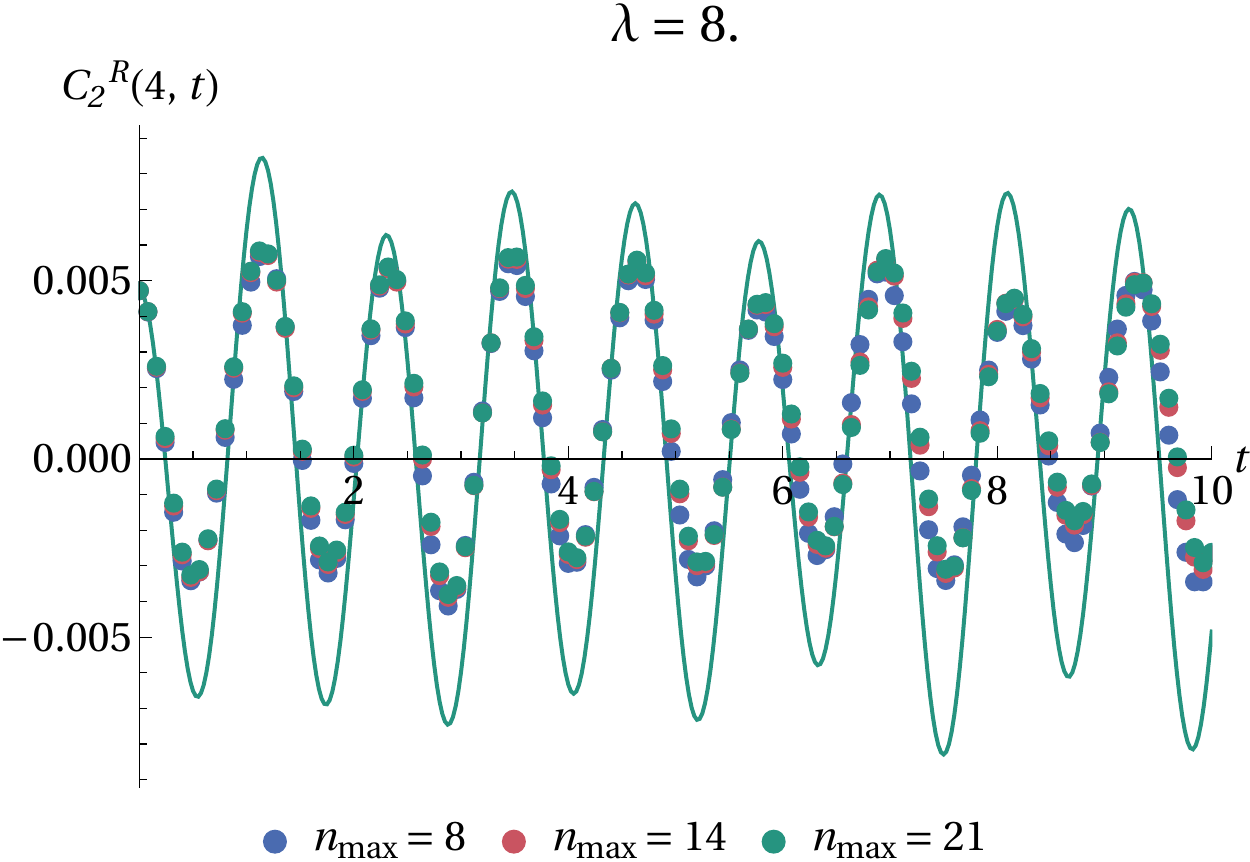}
\caption{Cutoff-dependence of $C_2(4,t)$. The dots of various colour show the THA results with different truncations parameterised by $n_\text{max}$ (using the renormalisation improvement), while the continuous line is the SCA result.}
\label{fig:C24truncationeffects}
\end{center}
\end{figure}

Note that despite the good agreement of the temporal frequencies for higher modes there is still a discrepancy in the amplitudes of the post-quench oscillations. One possible origin for that could be truncation errors from THA, since the method is expected to give poorer results for quantities with higher frequencies, closer to the truncation energy $\Lambda$. However, this can be safely excluded by examining the cutoff dependence of $C_2(4,t)$, shown in Fig. \ref{fig:C24truncationeffects}, confirming that the cutoff dependence of THA is clearly insufficient to explain the discrepancy between the THA and SCA. This leaves us with the only possibility that the discrepancy stems from the mode-mode correlations neglected in the mean-field approximation \eqref{hf}.

\begin{figure*}
\begin{center}
\begin{subfigure}[b]{\textwidth}
\includegraphics[width=0.48\textwidth]{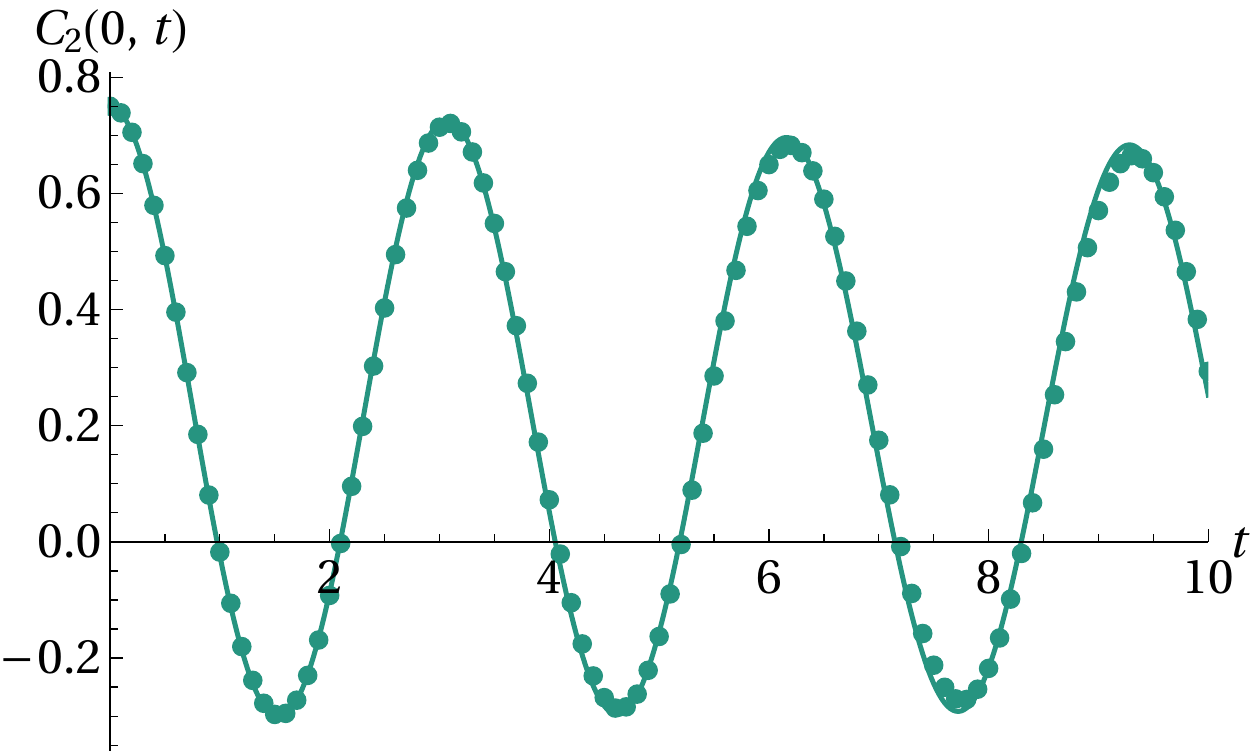}
\includegraphics[width=0.48\textwidth]{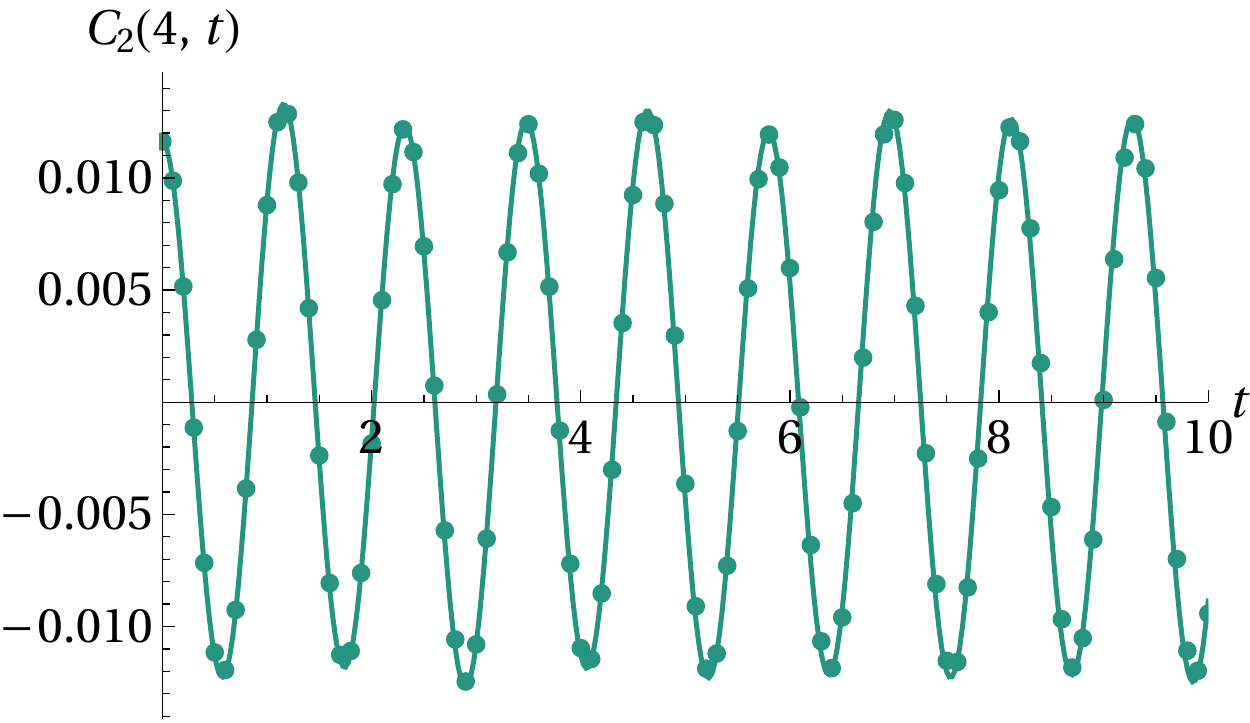}
\caption{$(m_0,\lambda_0) = (0.4,0) \rightarrow (m_1,\lambda_1) = (1,1)$}
\label{fig:c2-m4-lambda1}
\end{subfigure}
\begin{subfigure}[b]{\textwidth}
\includegraphics[width=0.48\textwidth]{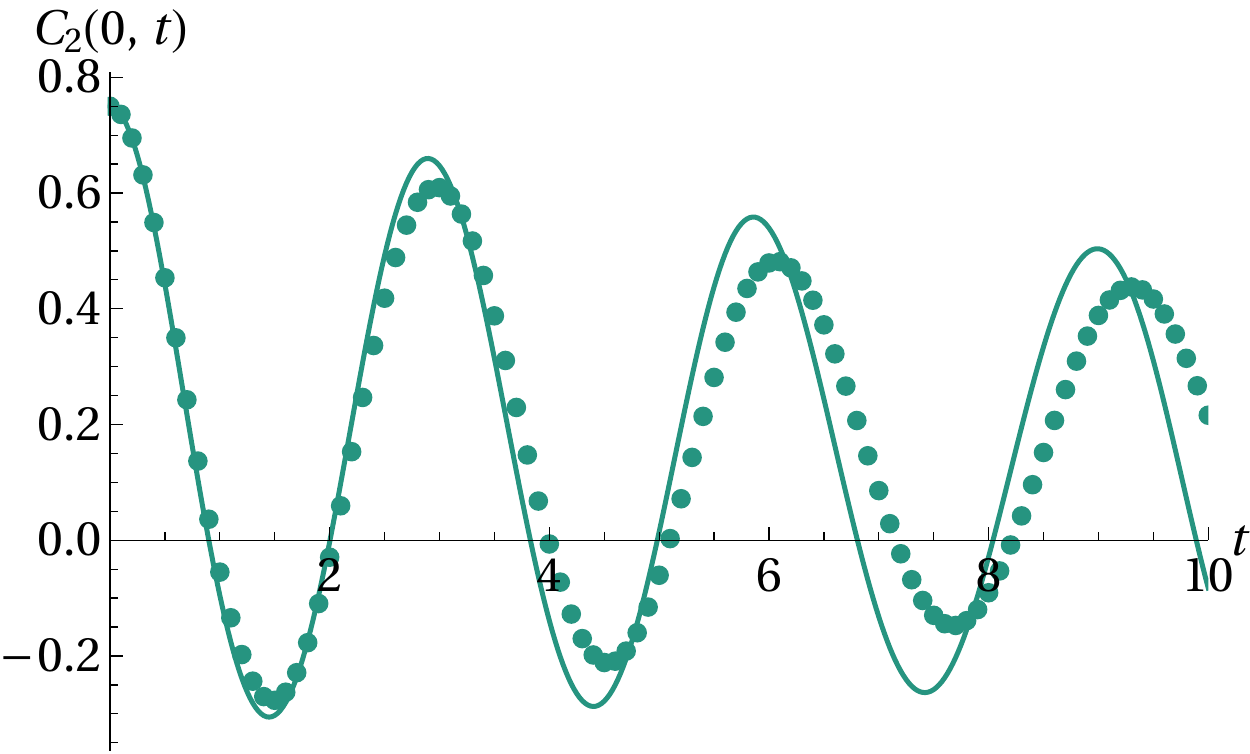}
\includegraphics[width=0.48\textwidth]{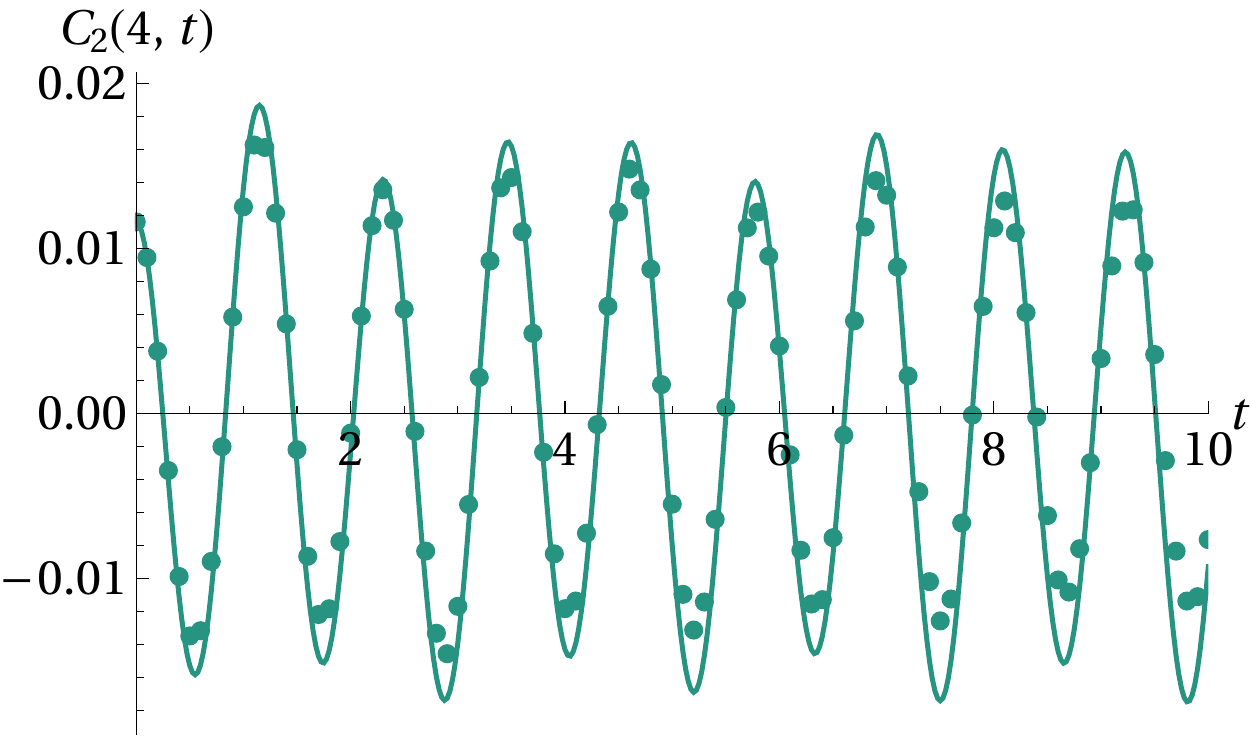}
\caption{$(m_0,\lambda_0) = (0.4,0) \rightarrow (m_1,\lambda_1) = (1,4)$}
\label{fig:c2-m4-lambda4}
\end{subfigure}
\begin{subfigure}[b]{\textwidth}
\includegraphics[width=0.48\textwidth]{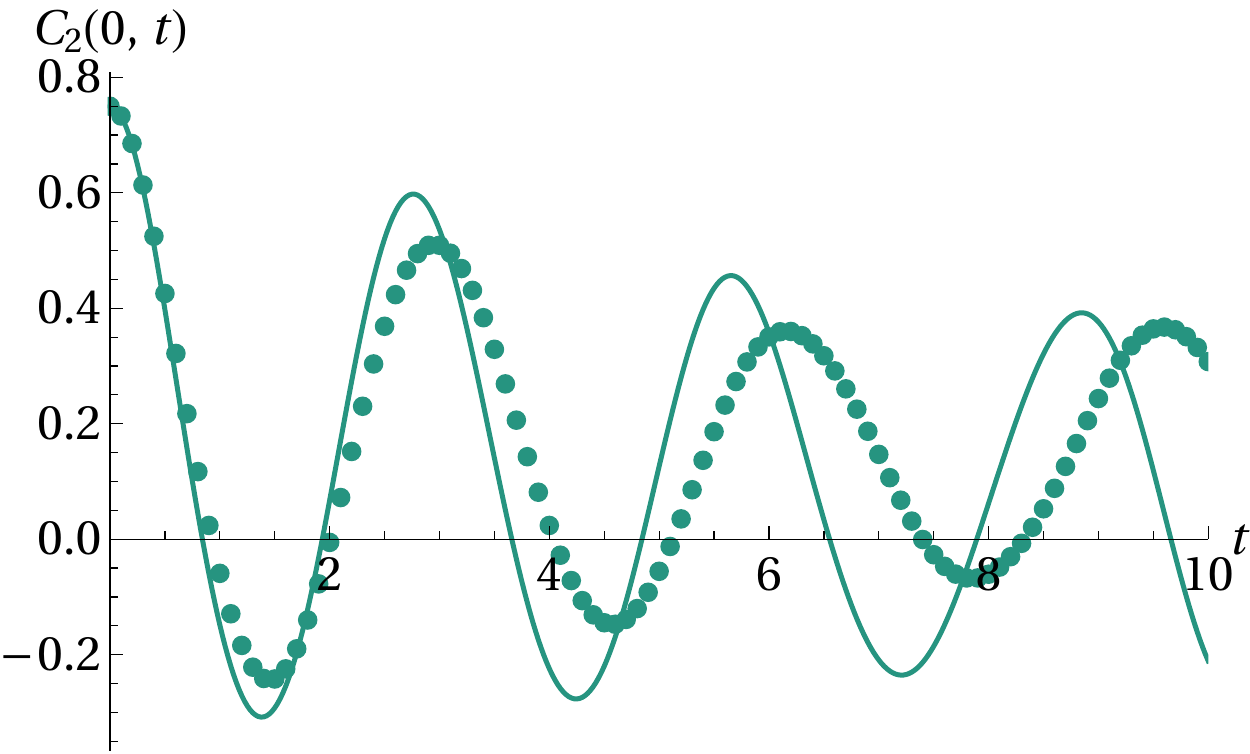}
\includegraphics[width=0.48\textwidth]{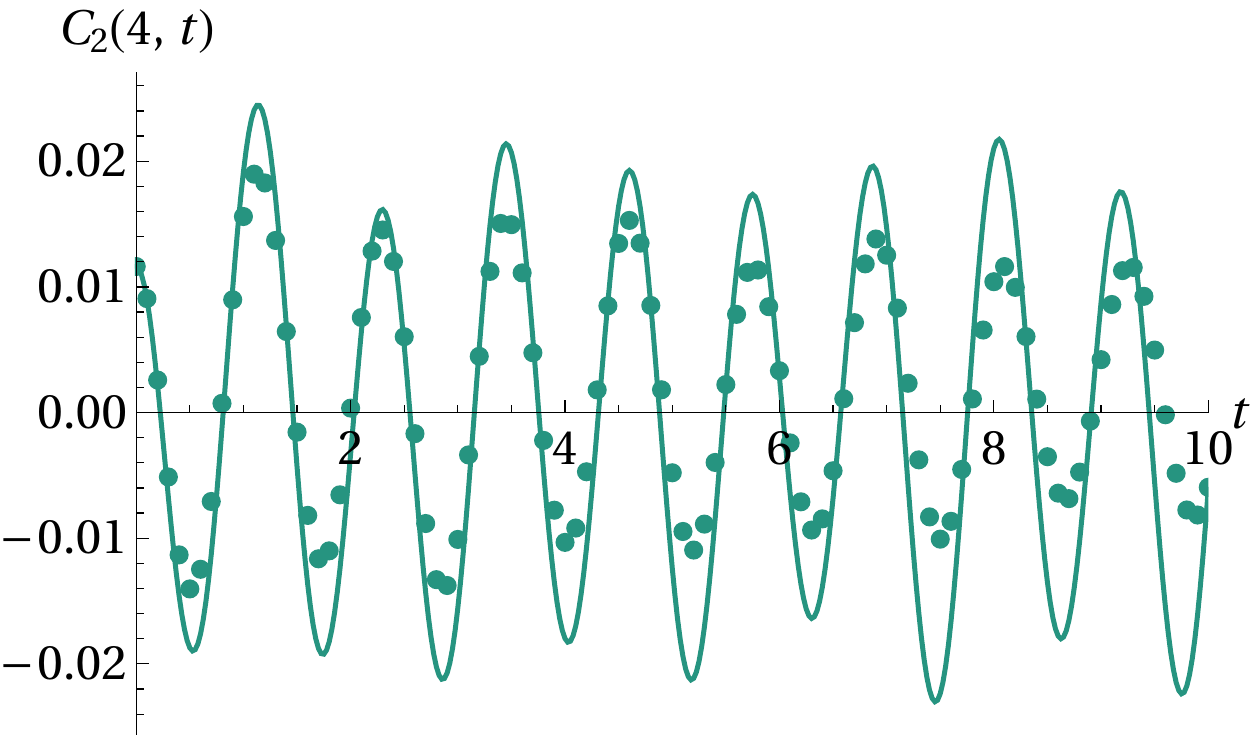}
\caption{$(m_0,\lambda_0) = (0.4,0) \rightarrow (m_1,\lambda_1) = (1,7)$}
\label{fig:c2-m4-lambda7}
\end{subfigure}
\caption{Time evolution of the $n = 0$ and $n = 4$ mode of $C_2$ for quenches with the mass changed from $0.4$ to $1$, with different post-quench interactions. The dots denote the numerical data from the truncated Hamiltonian approach at the cutoff  $n_\textrm{max}=21$ computed using the renormalisation improvement \eqref{eq:renormHam}, while the solid line is the result of the self-consistent approximation.}
\label{fig:c2-m4-lambdas}
\end{center}
\end{figure*}

The attribution of the failure of the SCA at strong coupling is further confirmed by considering a quench with a larger change in mass, shown in Fig. \ref{fig:c2-m4-lambdas}. Note that a larger change in mass leads to a larger quench in the sense that the post-quench energy density is higher, which in principle makes truncated Hamiltonian methods less precise \cite{Rakovszky_2016}. However, for the quenches shown in  Fig. \ref{fig:c2-m4-lambdas} the results for the highest truncation can still be considered identical to the exact quantum time evolution as discussed in Section \ref{sec:THA}. We recall that the SCA is exact for pure mass quenches $\lambda_0=\lambda_1=0$, and we also see that the discrepancy follows the same pattern in terms of the post-quench interaction parameter $\lambda_1$ as for the smaller quenches shown in Fig. \ref{fig:c2-lambdas}. 

\section{Truncated Wigner approximation vs. truncated Hamiltonian approach}\label{sec:twa}

Besides the self-consistent approximation discussed above, other semi-classical approaches can be applied to describe quantum quenches in interacting many-body systems~\cite{Polkovnikov2003,POLKOVNIKOV2010,2021EPJC...81..704R}. In particular, the truncated Wigner approximation (TWA) was argued to provide valuable insight into the dynamics of interacting field theories, such as the sine-Gordon model~\cite{2019PhRvA.100a3613H}. However, the accuracy of the TWA is hard to control, warranting a careful validation of the approximation when it is applied to a different Hamiltonian or quench procedure. Here we test TWA against THA for quenches in the $\phi^4$  model. First in Sec.~\ref{subsec:lattreg} we rewrite the $\phi^4$ theory as a lattice regularised model that serves as a convenient starting point for the TWA. Then in Sec.~\ref{subsec:twadetails} we review the details of TWA, and proceed to present the numerical results for mass and coupling quenches in Sec.~\ref{subsec:twanumerics}.

\subsection{Lattice regularisation}\label{subsec:lattreg}

In order to formulate TWA it is convenient to pass to the lattice regularised version of the $\phi^4$ model:
\begin{equation}
 H_{\mathrm{latt}}=\sum_{j=1}^{N_{s}}\left(\frac{\Pi_{j}^{2}+(\phi_j-\phi_{j-1})^2}{2a}
+\dfrac{m^2 a}{2}\,\phi_j^2 
+\dfrac{\lambda a}{4!}:\phi_j^4:\right)\,.
\label{eq:lattreg}
\end{equation}
Here $N_s$ denotes the number of lattice sites, $a=L/N_s$ is the lattice constant, $\Pi_j=\partial_t\phi_j / a$, and the operators $\phi_j$ and $\Pi_j$ satisfy the canonical commutation relation
\begin{equation}
    [\Pi_j,\phi_{j\prime}]=-i\delta_{jj\prime}\,.
\end{equation}
As above, $:\phi_j^4:$ stands for the normal ordered operator with respect to the post-quench mass. The spectrum of the quadratic part of $H_{\mathrm{latt}}$ can be obtained by setting $\lambda=0$, and diagonalising the Hamiltonian via the Fourier transform 
\begin{equation}
    \phi_k=\frac{1}{\sqrt{N_s}}\sum_je^{-ijak}\phi_j\quad
    \Pi_k=\frac{1}{\sqrt{N_s}}\sum_je^{-ijak}\Pi_j\,.
\end{equation} 
The resulting lattice dispersion
\begin{equation}\label{eq:lattdispersion}
    \omega_{k}^{\rm latt}=\sqrt{\left(\frac{2}{a}\sin\frac{ka}{2}\right)^2+m^2}
\end{equation}
reduces to the continuum relation $\omega_k= \sqrt{k^2+m^2}$ in the limit of small wave numbers $k\ll 1/a$. The vacuum correlators of the operators $\phi$ and $\Pi$ are given by
\begin{align}\label{eq:gscorr}
&\langle\phi_k\phi_{k\prime} \rangle_{\rm vac}=\delta_{k,-k^\prime}\,\dfrac{1}{2a\,\omega_{k}^{\rm latt}},\nonumber\\ &\langle\Pi_k\Pi_{k\prime} \rangle_{\rm vac}=\delta_{k,-k^\prime}\,\dfrac{a\,\omega_{k}^{\rm latt}}{2}.
\end{align}
The semi-classical TWA is written in terms of operators that are not normal ordered. Therefore, to treat $H_{\mathrm{latt}}$ semi-classically, $:\phi_j^4:$ must be rewritten as
\begin{equation*}
    :\phi_j^4:=\phi_j^4-\phi_j^2\,\dfrac{\lambda a}{4 N_s}\sum_k\dfrac{1}{2a\,\omega_k^{\rm latt}}+{\rm const.},
\end{equation*}
with the sum running over the wave numbers $k=2n\pi/L$, $n=0,\pm 1,\pm 2,...,N_s/2$, and the second term accounting for the vacuum expectation value $\langle\phi_k\phi_{-k} \rangle_{\rm vac}$. Defining the bare mass
\begin{equation}\label{eq:mbare}
    m_{\rm bare}^2 = m^2-\dfrac{\lambda}{4N_s a}\sum_k\dfrac{1}{\omega_k^{\rm latt}},
\end{equation}
the Hamiltonian ~\eqref{eq:lattreg} can be recast as
\begin{equation}\label{eq:Hlatt}
  H_{\mathrm{latt}}=\sum_{j=1}^{N_{s}}\left(\frac{\Pi_{j}^{2}+(\phi_j-\phi_{j-1})^2}{2a}+\dfrac{m_{\rm bare}^2 a}{2}\,\phi_j^2+\dfrac{\lambda a}{4!}\phi_j^4\right).
\end{equation}

\subsection{Truncated Wigner approximation}\label{subsec:twadetails}

The semi-classical TWA approximates out-of-equilibrium expectation values and correlations as phase space averages over random classical trajectories~\cite{Polkovnikov2003,POLKOVNIKOV2010}. It is closely related to the standard mean field approximation; both approaches approximate the time evolution by solving the classical equations of motion for the phase space coordinates $\lbrace\phi_j,\Pi_j\rbrace$. However, TWA goes significantly beyond the mean field approximation by incorporating the quantum fluctuations of the initial state.  This is achieved by sampling random phase space coordinates, $\lbrace\phi_j^{(0)},\Pi_j^{(0)}\rbrace$, at time $t=0$, and by determining the classical trajectories for these fluctuating initial conditions. More formally, TWA can be constructed through a systematic expansion of the Keldysh path integral~\cite{Polkovnikov2003,POLKOVNIKOV2010}. 

In this section we summarise the TWA for Hamiltonian ~\eqref{eq:Hlatt}. First we describe the general formulas allowing the calculate the time dependent expectation value $\langle \mathcal{O}\rangle(t)$ of an arbitrary operator $\mathcal{O}$ for an initial state given by a density matrix $\rho_0$, and then we turn to formulating the TWA approximation for the quenches discussed in the previous sections. 

Focusing on the lattice Hamiltonian~\eqref{eq:Hlatt}, it is convenient to  introduce the notations $|\phi\rangle_{j}$
and $|\Pi\rangle_{j}$ for the eigenstates of the operators $\phi_{j}$
and $\Pi_{j}$, which satisfy the completeness relations 
\begin{equation}
{\rm \mathbb{I}_{j}}=\int\dfrac{{\rm d}\Pi}{2\pi}\,|\Pi\rangle_j\,_j\langle\Pi|=\int{\rm d}\phi\,|\phi\rangle_j\,_j\langle\phi|
\end{equation}
at any site $j$, while their overlap is given by 
\begin{equation}
_{j}\langle\phi|\Pi\rangle_{j}=e^{i\phi \Pi}\,.\label{eq:overlap}
\end{equation}
We introduce a more compact vector notation 
\begin{equation}
\underline{\phi}=\lbrace\phi_{j}\,|j=1,...,N_{s}\rbrace
\end{equation}
for the full set of eigenvalues, with analogous notation for the eigenvalues
of the canonical conjugate operators $\Pi_{j}$.

The initial state characterised by the density matrix $\rho_0$ can be described by a quasi-probability distribution in phase space, given by the Wigner function
\begin{equation}
    W(\underline{\phi},\underline{\Pi})=\dfrac{1}{(2\pi)^{2N_{s}}}\int{\rm d}\underline{\phi}^{\prime}\,\langle\underline{\phi}+\underline{\phi}^{\prime}/2|\,\rho_0\,|\underline{\phi}-\underline{\phi}^{\prime}/2\rangle\,e^{-i\underline{\phi}^{\prime}\underline{\Pi}}.\label{eq:W}
\end{equation}
Similarly, an arbitrary operator $\mathcal{O}$ can be represented as a function over the phase space coordinates, by its Wigner transform defined as
\begin{align}
O_{W}(\underline{\phi},\underline{\Pi})=\int{\rm d}\underline{\phi}^{\prime}\,\langle\underline{\phi}-\underline{\phi}^{\prime}/2|\,\mathcal{O}\,|\underline{\phi}+\underline{\phi}^{\prime}/2\rangle\,e^{i\underline{\phi}^{\prime}\underline{\Pi}}.\label{eq:Ow}
\end{align}
The TWA can then be formulated in terms of the phase space distribution $W$ and phase space functions $O_W$. The quantum fluctuations of the initial state are incorporated by sampling random initial conditions, $(\underline{\phi}^{(0)},\underline{\Pi}^{(0)})$, from the Wigner quasi-distribution. The time evolution is then obtained by solving the classical equations of motion,
\begin{align}
 & \partial_{t}\Pi_{j}=-\dfrac{1}{ a}(\phi_{j+1}+\phi_{j-1}-2\phi_{j})-m_{\rm bare}^2 a\,\phi_j-\dfrac{\lambda a}{6}\phi_{j}^3\,,\nonumber \\
 & \partial_{t}\phi_{j}=\dfrac{1}{a}\Pi_{j}\,,\label{eq:EOM}
\end{align}
with the bare mass $m_{\rm bare}^2$ defined in Eq.~\eqref{eq:mbare}.
Given the trajectory $(\underline{\phi}(t),\underline{\Pi}(t))$, its contribution to the operator expectation value $\langle\mathcal{O}\rangle$ can be evaluated  by substituting the fields $\underline{\phi}(t)$ and $\underline{\Pi}(t)$ into the Wigner transform $O_W$. Finally, the TWA expectation value $\langle\mathcal{O}\rangle_{\mathrm{TW}}$ is obtained by averaging over a large number of different initial conditions $(\underline{\phi}^{(0)},\underline{\Pi}^{(0)})$.
The procedure outlined above can be summarised in a  compact form as 
\begin{eqnarray}
&&\langle\mathcal{O}\rangle_{\mathrm{TW}}(t)=\nonumber\\
&&\int\!\!\!\int\!{\rm d}\underline{\phi}^{(0)}{\rm d}\underline{\Pi}^{(0)}\,W(\underline{\phi}^{(0)},\underline{\Pi}^{(0)})\,O_{W}(\underline{\phi}(t),\underline{\Pi}(t))\,,\label{eq:TW}
\end{eqnarray}
expressing the operator expectation value as a phase space average over random classical trajectories, weighted according to the Wigner quasi-distribution of the initial state.

For the quenches under consideration the initial state is the ground state of a free boson model with $\lambda=0$, described by a simple Gaussian Wigner function
\begin{align}\label{eq:Wquench}
    W=&\prod_{k=0,\pi/a}\dfrac{1}{\pi}\exp\left(-\dfrac{\phi_{k}^2}{2\sigma_{k}^{2}}-2\sigma_{k}^{2}\,\Pi_{k}^2\right)\cdot\nonumber\\
    &\prod_{0<k<\pi/a}\dfrac{4}{\pi^{2}}\exp\left(-\dfrac{\phi_{k}\,\phi_{-k}}{\sigma_{k}^{2}}-4\sigma_{k}^{2}\,\Pi_{k}\,\Pi_{-k}\right)\,.
\end{align}
Here $\sigma_k$ denotes the variance extracted from Eq.~\eqref{eq:gscorr},
\begin{equation*}
    \sigma_k=\dfrac{1}{2\,a\,\omega_{k,0}^{\rm latt}}\,,
\end{equation*}
where $\omega_{k,0}^{\rm latt}$ is given by Eq.~\eqref{eq:lattdispersion}, calculated with the initial renormalised mass $m_0$. 

We focus on the time evolution of the correlation function
\begin{equation*}
    C_2(k,t)=a\langle :\phi_k(t)\phi_{-k}(t):\rangle=a\langle \phi_k(t)\phi_{-k}(t)\rangle-\dfrac{1}{2\,\omega_{k,1}^{\rm latt}},
\end{equation*}
with $\omega_{k,1}^{\rm latt}$ given by Eq.~\eqref{eq:lattdispersion} at the post-quench renormalised mass $m_1$. Here the lattice constant $a$ is inserted to match the normalisation of the lattice model and the continuum theory. Since $C_2$ only depends on the phase $\phi$, the Wigner transform~\eqref{eq:Ow} becomes trivial,  and amounts to substituting operators with classical variables:
\begin{equation}\label{eq:C2w}
    C_{2,W}(k,t)=a \,\phi_k(t)\phi_{-k}(t)-\dfrac{1}{2\,\omega_{k,1}^{\rm latt}}\,.
\end{equation}
The TWA simulations proceed as follows. We generate the initial  conditions $(\underline{\phi}^{(0)},\underline{\Pi}^{(0)})$ from the Gaussian distribution ~\eqref{eq:Wquench}, then calculate the time evolution from ~\eqref{eq:EOM}. The contribution of a given trajectory to the correlator $C_2$ is evaluated from Eq.~\eqref{eq:C2w}. Finally, we perform an averaging over the initial conditions. Our findings are presented in the next subsection. 

\subsection{Quenches in the $\phi^4$ theory: TWA vs. THA}\label{subsec:twanumerics}

\begin{figure*}
\centering
\begin{subfigure}[b]{0.48\textwidth}
\centering
\includegraphics[width=\textwidth]{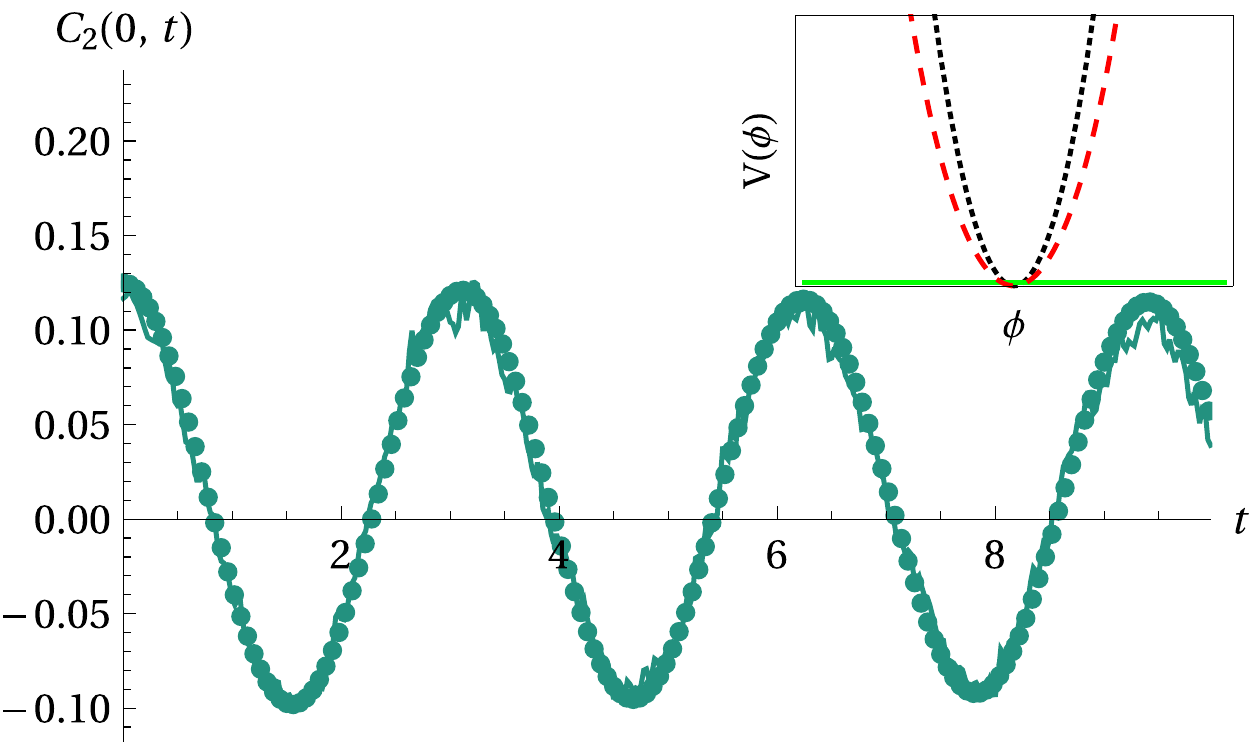}
\caption{$(m_0,\lambda_0) = (0.8,0) \rightarrow (m_1,\lambda_1) = (1,1)$}
\label{fig:twa-m00_8-l1}
\end{subfigure}
\centering
\begin{subfigure}[b]{0.48\textwidth}
\includegraphics[width=\textwidth]{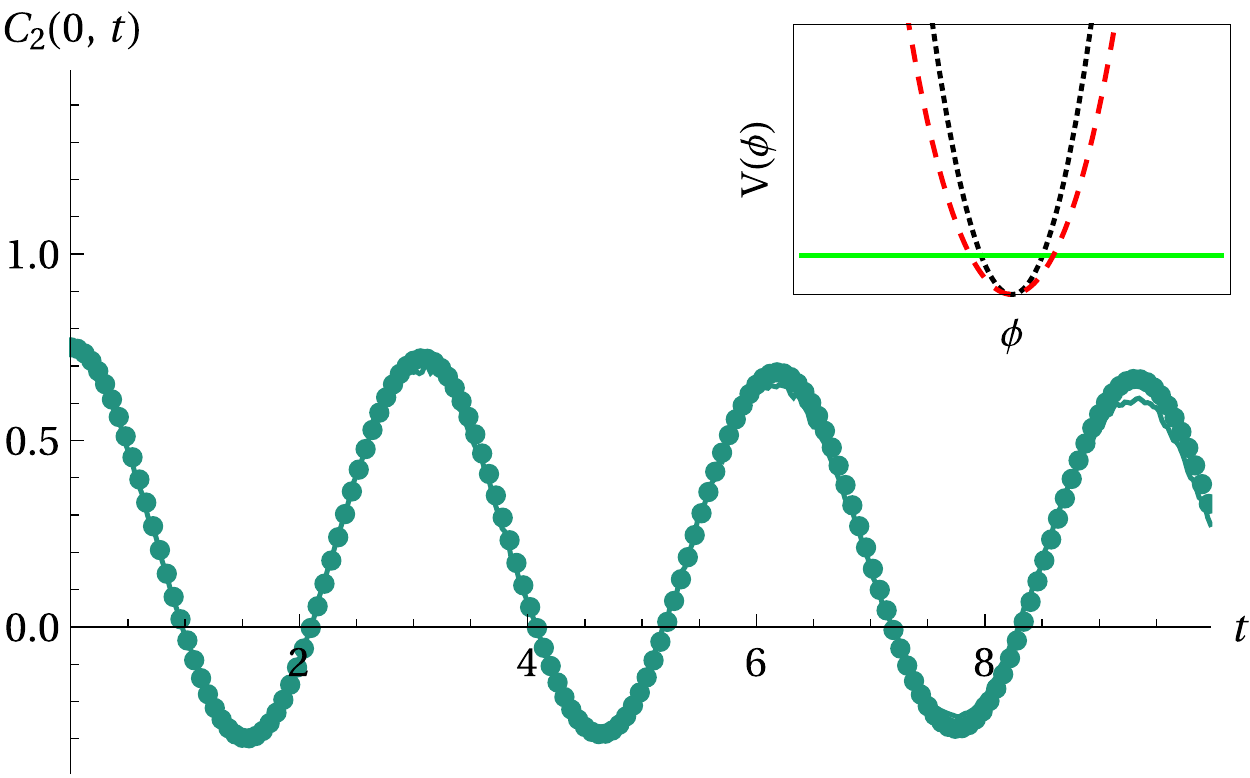}
\caption{$(m_0,\lambda_0) = (0.4,0) \rightarrow (m_1,\lambda_1) = (1,1)$}
\label{fig:twa-m00_4-l1}
\end{subfigure}
\\
\begin{subfigure}[b]{0.48\textwidth}
\centering
\includegraphics[width=\textwidth]{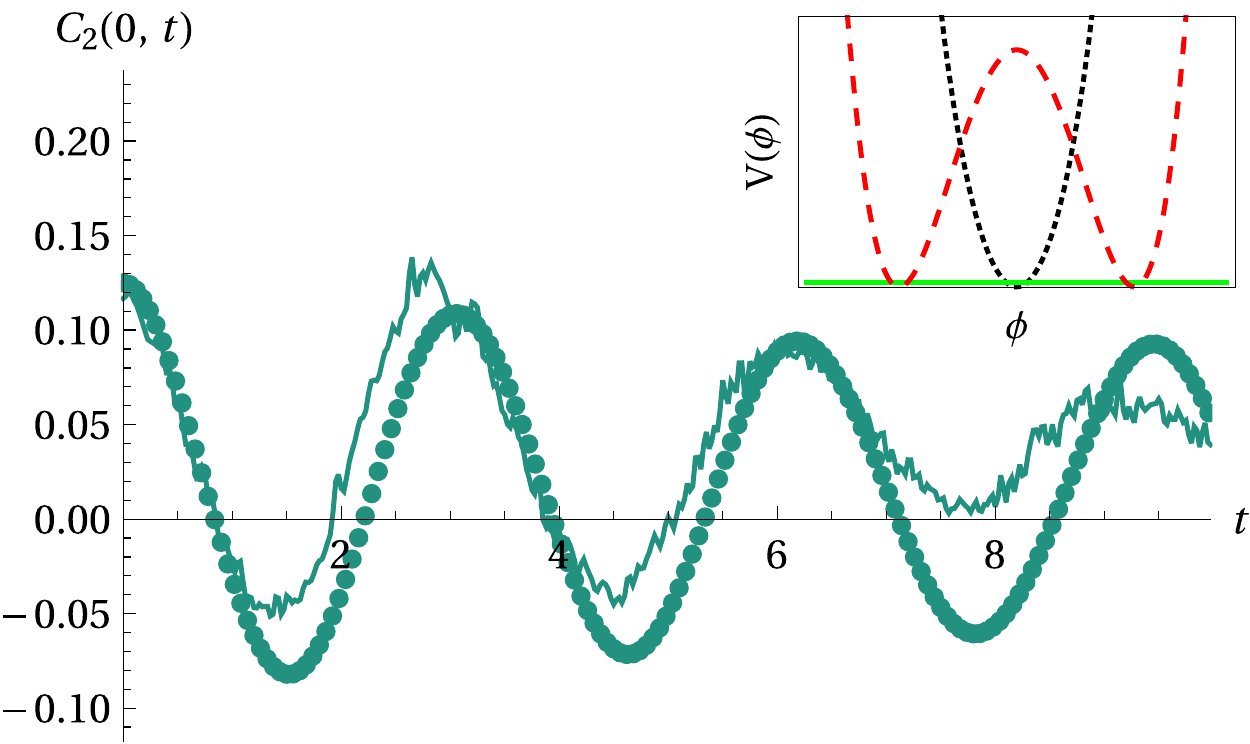}
\caption{$(m_0,\lambda_0) = (0.8,0) \rightarrow (m_1,\lambda_1) = (1,4)$}
\label{fig:twa-m00_8-l4}
\end{subfigure}
\centering
\begin{subfigure}[b]{0.48\textwidth}
\includegraphics[width=\textwidth]{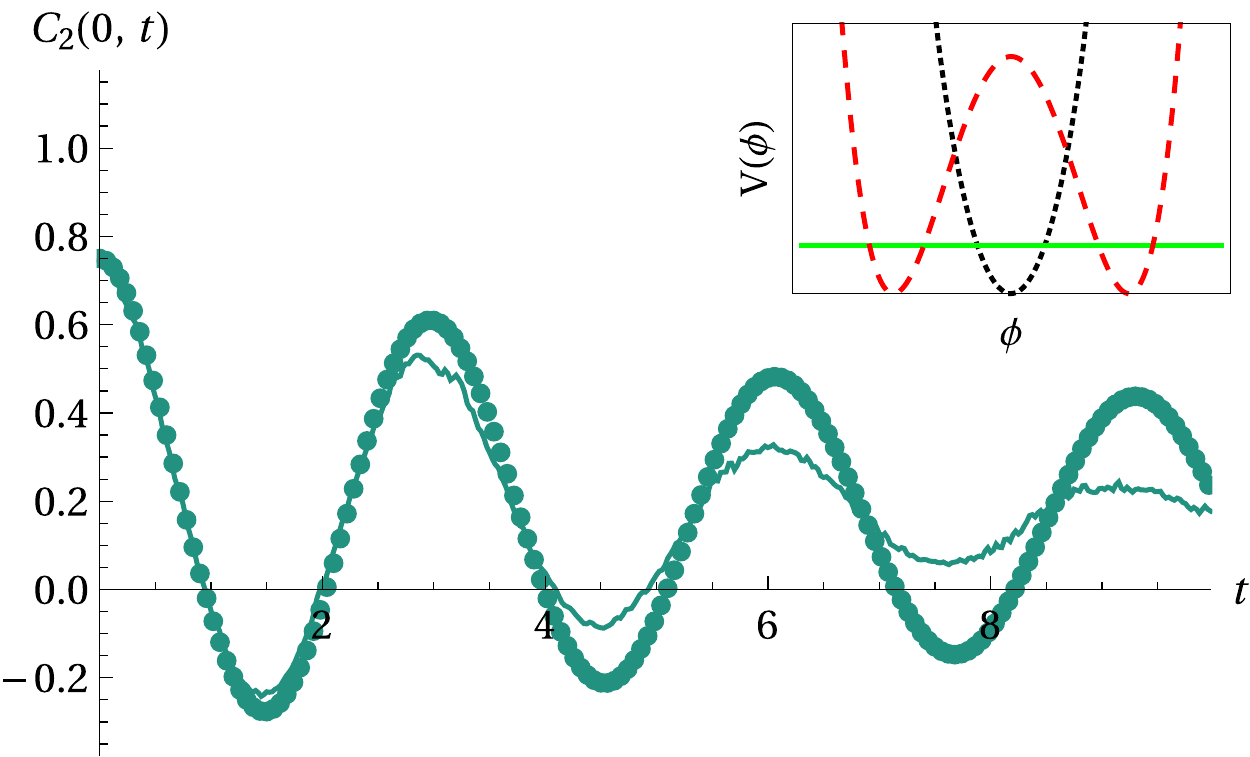}
\caption{$(m_0,\lambda_0) = (0.4,0) \rightarrow (m_1,\lambda_1) = (1,4)$}
\label{fig:twa-m00_4-l4}
\end{subfigure}
\\
\begin{subfigure}[b]{0.48\textwidth}
\centering
\includegraphics[width=\textwidth]{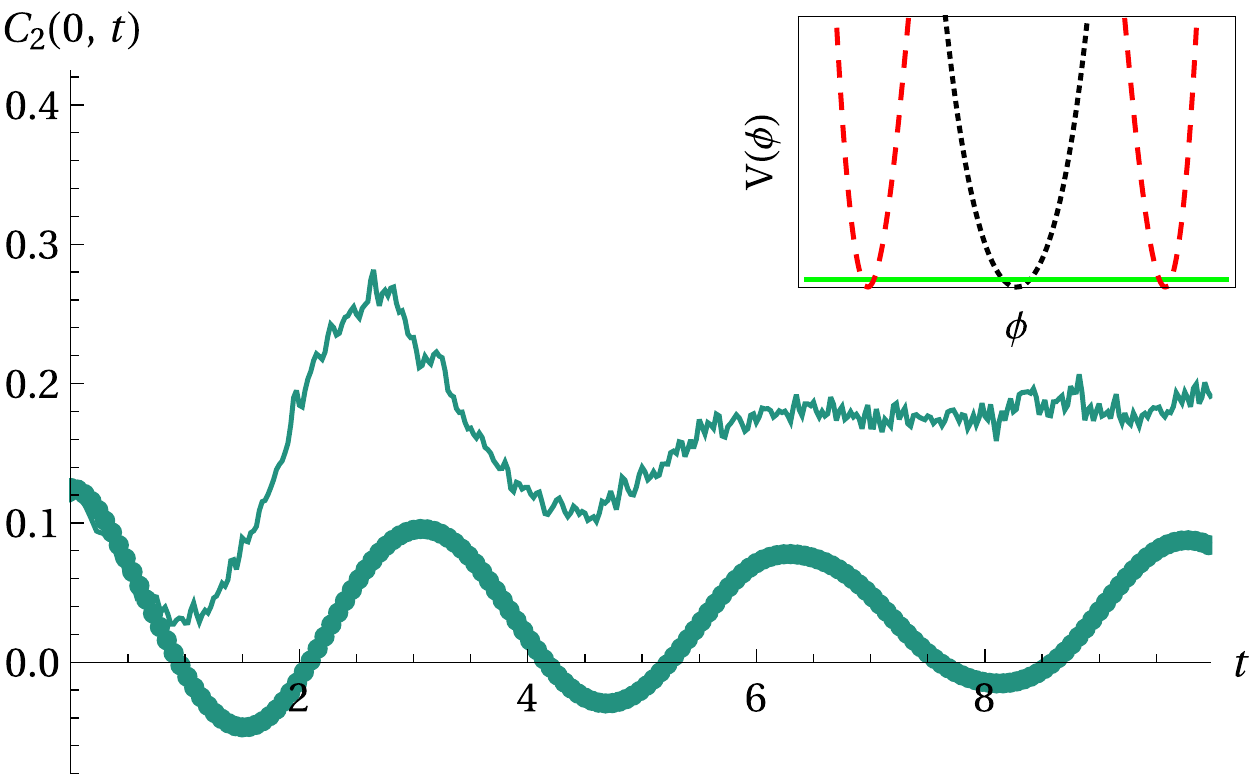}
\caption{$(m_0,\lambda_0) = (0.8,0) \rightarrow (m_1,\lambda_1) = (1,8)$}
\label{fig:twa-m00_8-l8}
\end{subfigure}
\centering
\begin{subfigure}[b]{0.48\textwidth}
\includegraphics[width=\textwidth]{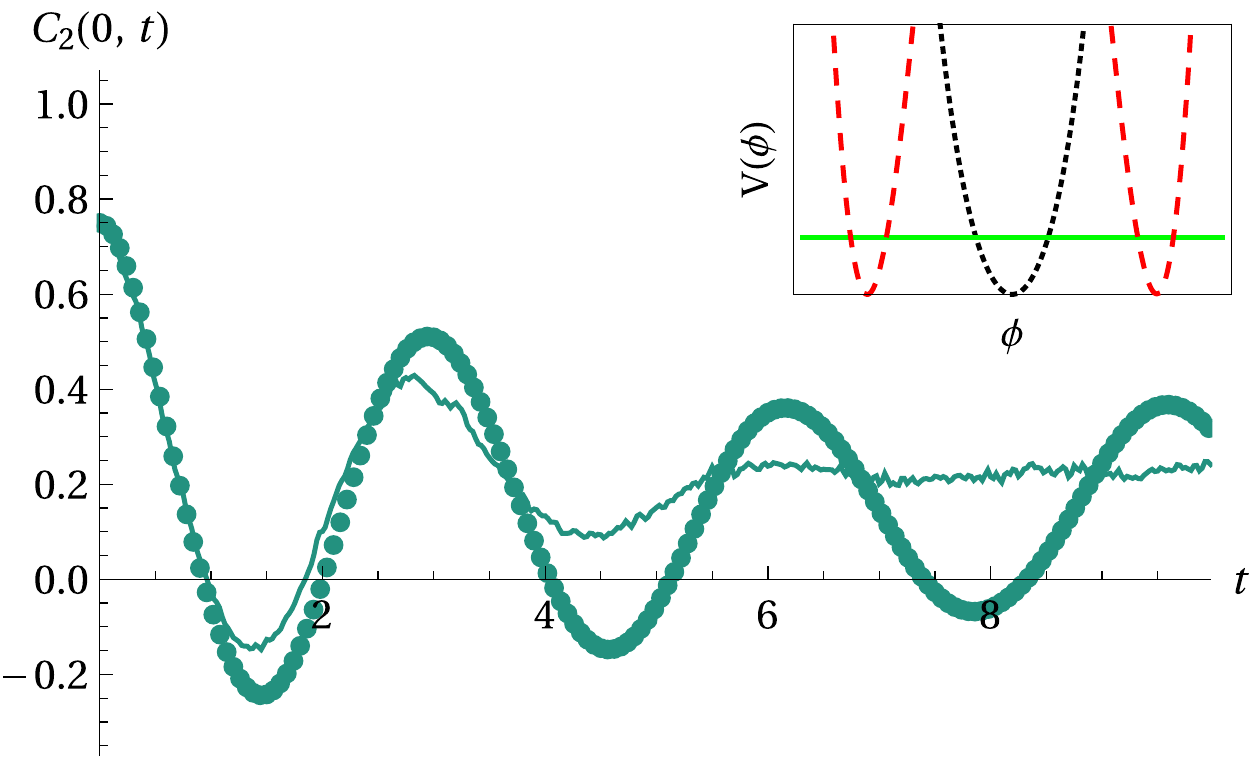}
\caption{$(m_0,\lambda_0) = (0.4,0) \rightarrow (m_1,\lambda_1) = (1,7)$}
\label{fig:twa-m00_4-l7}
\end{subfigure}
\caption{Time evolution of the correlator $C_2(k=0,t)$, calculated from TWA (solid line) and THA (dots), for a quenches to weak/intermediate/strong interactions, with two different values for the mass shift. We used $N_s=400$ in TWA, and cutoff $n_{\rm max}=21$ (including the leading RG improvement) in THA. {The insets compare the THA potential (dotted black line), the TWA potential (dashed red line) and the energy density injected by the quench (green solid line) computed as $\Delta \mathcal{E}=( \braket{\psi_0|H_1|\psi_0} - E_0^{(1)})/L$ where $E_0^{(1)}$ is the vacuum energy of $H_1$.}}
 \label{fig:twa-tha}
\end{figure*}

Here we present our numerical results for the mass and interaction quenches of the $\phi^4$ theory and compare them to THA simulations, which can be considered to reflect the exact quantum time evolution due to their high precision, as discussed in Section \ref{sec:THA}.

Importantly, for pure mass quenches with $\lambda\equiv 0$, TWA reproduces the exact time evolution, similarly to the SCA discussed in Section \ref{sec:SC}. As shown in Fig.~\ref{fig:twa-m00_8-l1}, displaying the $C_2$ correlator for mode 0 for a small quench $m_0 = 0.8, \lambda_0 = 0 \rightarrow m_1 = 1, \lambda_1 = 1$, we continue to find a good agreement between TWA and THA slightly away from the free boson limit, therefore TWA remains valid for quenches into the interacting $\phi^4$ theory for small enough interaction strength $\lambda$. However, for larger $\lambda$ significant deviations develop between TWA and THA, pointing to the failure of TWA. In contrast to the more reliable THA, TWA predicts strongly damped oscillations in $C_2$, and the correlator for mode 0 relaxes to a finite stationary value. The relaxation rate, as well as the stationary limit of $C_2$, increases with $\lambda$, as illustrated in Figs.~\ref{fig:twa-m00_8-l4} and ~\ref{fig:twa-m00_8-l8}. 

We find a similar behaviour for larger mass quenches $m_0 = 0.4, \lambda_0 = 0 \rightarrow m_1 = 1, \lambda_1$, as shown in Figs. ~\ref{fig:twa-m00_4-l1}, ~\ref{fig:twa-m00_4-l4} and ~\ref{fig:twa-m00_4-l7}, displaying $C_2$ for mode 0 for post-quench interaction strengths $\lambda_1=1$, $4$ and 7, respectively. The TWA systematically overestimates the damping rate of oscillations, and this effect becomes more pronounced as the interaction strength $\lambda_1$ increases.

The failure of TWA at large interactions $\lambda_1$ can be attributed to sensitivity of the time evolution to the mass renormalisation. Note that instead of working directly with the renormalised mass $m_1$, TWA is formulated in terms of the bare mass $m_{\rm bare}^2$, given by Eq.~\eqref{eq:mbare}. For large enough $\lambda_1$, $m_{\rm bare}^2$ becomes negative, corresponding to a symmetry broken classical steady state with $\phi_j^2\equiv-6\,m_{\rm bare}^2/\lambda_1$, {as can be seen in the insets of Fig. \ref{fig:twa-tha}}. The classical equations of motion, Eq.~\eqref{eq:EOM}, then predict a fast relaxation to a finite value for the correlator $C_2$. In contrast, the exact renormalised mass entering THA remains positive, giving rise to a much weaker damping of oscillations. {Besides the coupling $\lambda_1$, the failure of the TWA also depends on the energy density $\Delta \mathcal{E}$ injected in the quench, which can be controlled varying the pre-quench mass $m_0$ while keeping the post-quench parameters $m_1$ and $\lambda_1$ determining the shapes of the THA and TWA potentials fixed. As shown in the insets of Fig. \ref{fig:twa-tha}, the damping in the TWA and the deviation from THA is exacerbated by smaller values of $\Delta \mathcal{E}$ (corresponding to the larger values of $m_0$ on the left), since it makes the classical trajectories contributing to the TWA more strongly constrained in the neighbourhood of the ``fake'' symmetry breaking minima of the THA potential.}  

\begin{figure*}
\centering
\begin{subfigure}[b]{0.48\textwidth}
\centering
\includegraphics[width = \columnwidth]{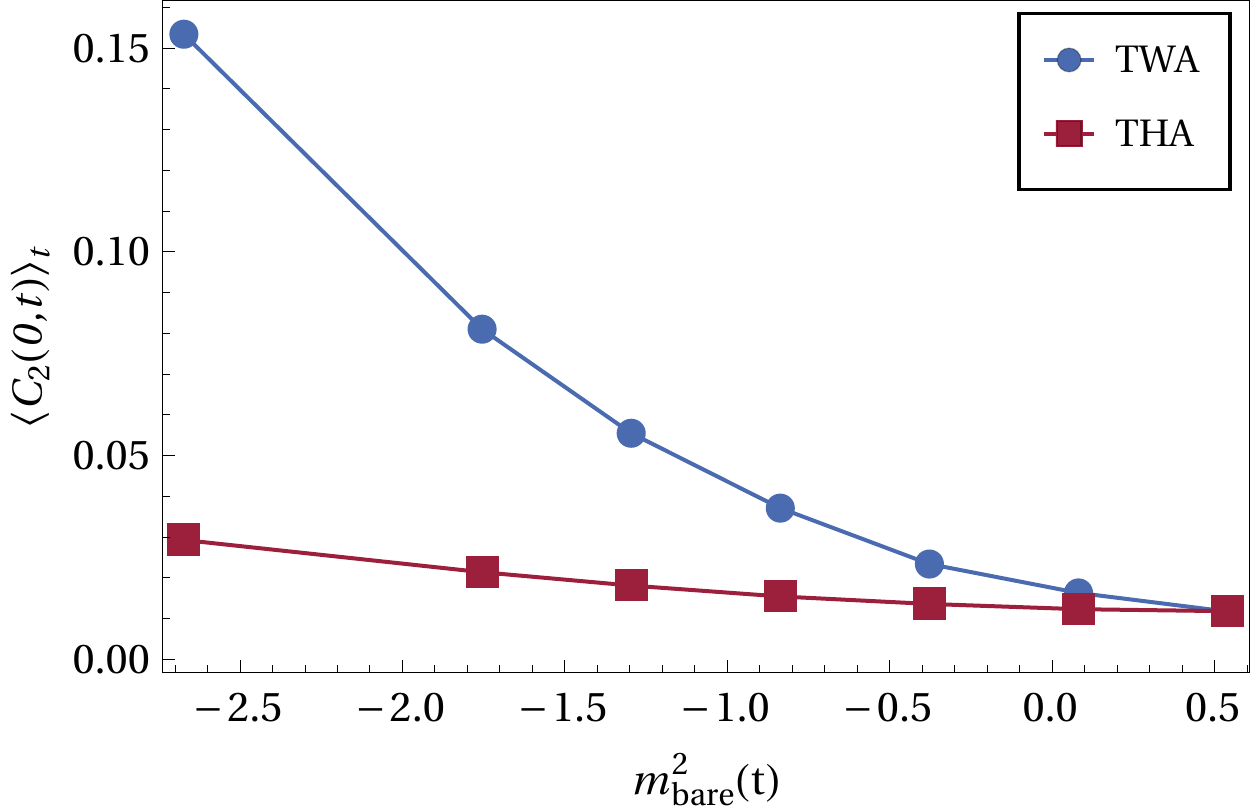}
\caption{Quenches $m_0 = 0.8, \lambda_0 = 0 \rightarrow m_1 = 1, \lambda_1$}
\label{fig:C2av_m00_8}
\end{subfigure}
\begin{subfigure}[b]{0.48\textwidth}
\centering
\includegraphics[width = \columnwidth]{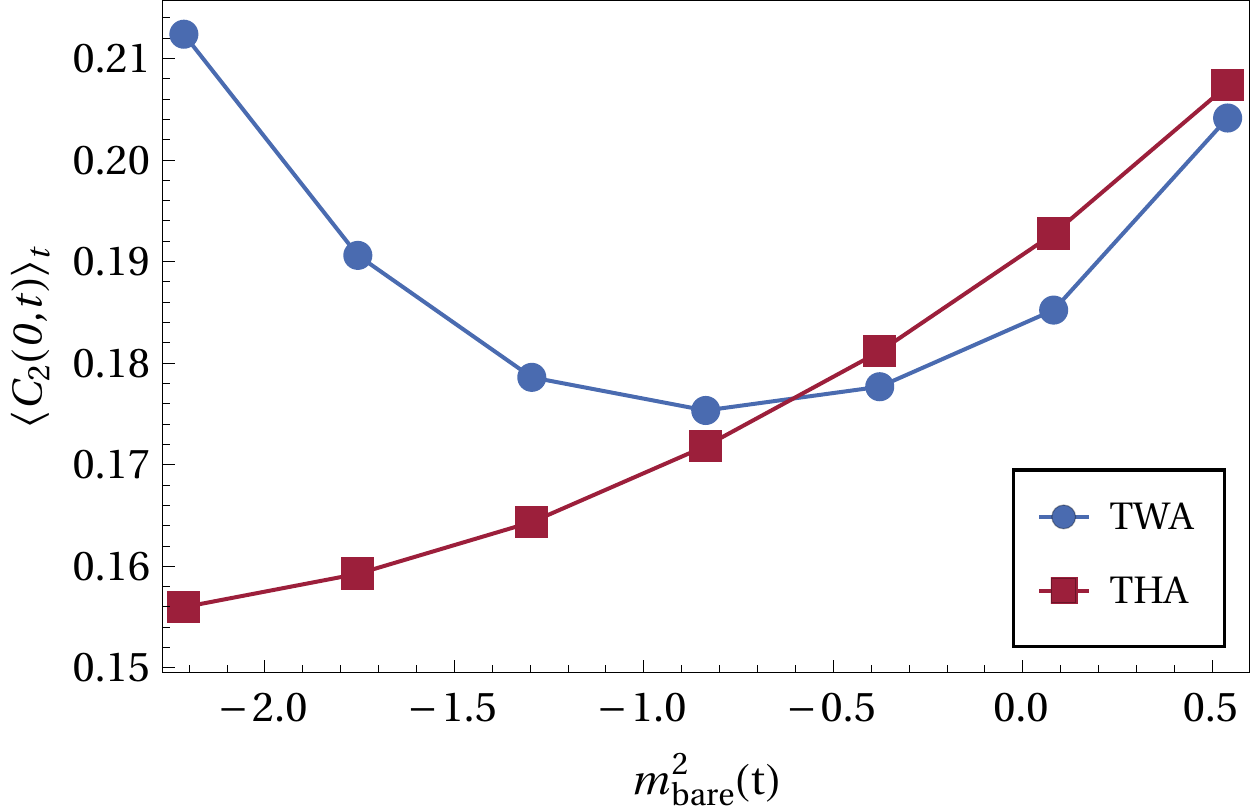}
\caption{Quenches $m_0 = 0.4, \lambda_0 = 0 \rightarrow m_1 = 1, \lambda_1$}
\label{fig:C2av_m00_4}
\end{subfigure}
\caption{Time averaged correlator $\langle C_2(k=0,t)\rangle_t$, calculated from TWA (blue circles) and THA (red squares), with different interactions $\lambda_1$, plotted as a function of the bare mass $m_{\rm bare}^2$ \eqref{eq:mbare}. The TWA and THA results show a good agreement for $m_{\rm bare}^2>0$, but start to deviate when it becomes negative with increasing $\lambda_1$, with TWA predicting a considerably higher average. We used $N_s=400$ in TWA, and cutoff $n_{\rm max}=21$ (including the leading RG improvement) in THA.}
\label{fig:C2av}
\end{figure*}

Further support for the responsibility of the negative bare mass for the enhanced damping in TWA compared to THA, can be obtained by considering the time averaged correlator
\begin{equation}
    \langle C_2(k,t)\rangle_t = \dfrac{1}{T}\int_0^T {\rm d}t\,C_2(k,t)\,,
\end{equation}
where $T$ is chosen as the time of the first three  full oscillations. We calculated $\langle C_2(k=0)\rangle_t$ both with TWA and THA for quenches with fixed $m_0$ and $m_1$, varying $\lambda_1$, with the results shown in Fig. \ref{fig:C2av} as a function of  $m_{\rm bare}^2$, determined from $\lambda_1$ through Eq.~\eqref{eq:mbare}. We find a good agreement between TWA and THA as long as  $m_{\rm bare}^2$ stays positive. However, for $m_{\rm bare}^2<0$ TWA deviates from THA, and predicts an average $\langle C_2(k=0)\rangle_t$ that increases rapidly with decreasing $m_{\rm bare}^2$. For the larger quenches Fig.~\ref{fig:C2av_m00_4}, the deviation is somewhat larger between THA and TWA even in the region $m_{\rm bare}^2\geq 0$. However, in this case TWA and THA even predict opposite tendencies for $\langle C_2(k=0)\rangle_t$ when $m_{\rm bare}^2$ is negative: in TWA, the average correlator crosses over to a rapid increase with decreasing $m_{\rm bare}^2$, whereas the THA result continues to decrease even further. These results strongly support the argument that TWA fails for strong interaction $\lambda_1$ due to the negative bare mass entering the classical equations of motion, and that the strong damping in the TWA time evolution is an artefact originating from the mass renormalisation.

\section{Conclusions}\label{sec:conclusions}

In this paper we studied the dynamics following composite quenches in the gap and interaction coupling in the $\mathbb{Z}_2$-symmetric phase of the $(1+1)d$ $\phi^4$ theory using three different approximations: the truncated Hamiltonian approach (THA), the self-consistent approximation (SCA) and the truncated Wigner approximation (TWA). The quenches considered start from the ground state of a free massive boson, which is then evolved by a Hamiltonian with a different mass and non-zero self-interaction.

The THA used here was built upon the Hilbert-space of a free massive boson with a mass equal to the post-quench value. We used a large range of cutoff values together with renormalisation group improvement. However, the method turned out to be so convergent that the results showed very little dependence on the cutoff, and for the time window of the simulation could be taken to reflect the exact quantum time evolution. We also note that the system showed a very slow relaxation, consistent with an observation made by Durnin et al.\cite{2020arXiv200411030D} that the $\phi^4$ interaction belongs to the class of perturbations with weak integrability breaking\cite{2020arXiv200411030D,2021ScPP...11...37S}. 

The self-consistent approximation (SCA) to the non-equilibrium dynamics for the $\phi^4$ model was developed by Sotiriadis and Cardy\cite{Sotiriadis_2010}, and more recently applied to the dynamics of sine-Gordon quantum field theory \cite{2019JSMTE..08.4012V} and  tunnel-coupled one-dimensional Bose gases \cite{2020ScPP....9...25V}. It consists in a mean-field approximation to the interaction term, which reduces it to a time-dependent effective mass which can be self-consistently determined from a gap equation. It is expected to fail as the value of the self-interaction $\lambda$ increases, since it neglects the connected part of the four-point correlations which become more important. This is indeed what is observed in our calculations, with the only slight surprise that it happens for relatively small values of $\lambda$. The failure of the SCA shows up both in temporal frequencies for which the deviation is smaller for modes with larger wave numbers (faster spatial variation), and also in the amplitudes for which the deviation does not show any obvious dependence on the wave number $k$. These observations may be relevant in the light of recent applications of the self-consistent approximation\cite{2019JSMTE..08.4012V,2020ScPP....9...25V} mentioned above, albeit these works treated models different from the $\phi^4$ QFT considered here.

The truncated Wigner approach approximates out-of-equilibrium dynamics by solving the classical equations of motion with the quantum fluctuations incorporated in the initial state~\cite{Polkovnikov2003,POLKOVNIKOV2010}. The method, formulated for the dynamics of the lattice regularisation of the theory, contains the bare mass in its Hamiltonian which is renormalised by quantum fluctuations to become the physical mass. Comparison to the THA reveals that the TWA fails to reproduce the quantum time evolution whenever the bare mass becomes negative, giving rise to a symmetry broken steady state in the classical equations of motion. This effect can lead to a strong overestimation of the relaxation rate. This limitation of TWA is expected to be relevant for other models with a symmetry broken phase. 

The TWA has also been applied to the sine-Gordon model\cite{2013PhRvL.110i0404D}, and more recently\cite{2019PhRvA.100a3613H} its results were cross-checked against the truncated conformal space approach. It is interesting that for the sine-Gordon theory the problem with TWA observed here is entirely absent. The reason is that renormalisation in the sine-Gordon case works very differently from the case of the $\phi^4$ model. The sine-Gordon Hamiltonian is given by
\begin{equation}
    H_\textrm{sG}=\int dx \left(\frac{1}{2}\Pi^2+\frac{1}{2}(\partial_x\phi)^2-\lambda:\cos\beta\phi:\right)
\end{equation}
and renormalisation results from the normal ordering :: with respect to the modes of the massless boson described by the first two terms, which leads to a multiplicative renormalisation of the coupling $\lambda$\cite{2019PhRvA.100a3613H}, instead of the additive mass shift \eqref{eq:mbare} present in the $\phi^4$ model. As a result, the sign of the mass terms is never changed and the TWA is stable against the effects of renormalisation.

In contrast to the two semi-classical approaches, the THA performs well even in the strongly interacting regime, showing little truncation effects and yielding practically exact results for large enough values of the cutoff.  Therefore the THA is a very powerful and highly accurate method for studying time-evolution following quantum quenches in interacting field theories with relevant perturbations, especially the $\phi^4$ model.

We close by mentioning that since the self-consistent approximation is formulated as a semi-classical approximation in quantum field theory\cite{2021EPJC...81..704R}, a natural improvement is taking into account quantum effects using an approach such as the 2PI effective action\cite{2004AIPC..739....3B,2015PhRvX...5d1005B}, which is a promising direction for future investigations, for which the THA developed here provides an efficient source of validation.

\vspace{2mm}
\begin{acknowledgments}
This research was partially supported by by National Research, Development and Innovation Office (NKFIH) through the OTKA Grant K 138606, and also via the grant TKP2020 IES Grant No. BME-IE-NAT. G. T. was also supported by NKFIH via the Hungarian Quantum Technology National Excellence Program, project no. 2017-1.2.1-NKP- 2017-00001. I. L. acknowledges support from the European Research Council (ERC) under the European Union’s Horizon 2020 research and innovation programme (grant  agreement No. 771537).
\end{acknowledgments}

\bibliography{phi4.bib}

\end{document}